\documentclass[12pt, epsfig]{article}
\usepackage{amsmath,amsfonts,amssymb,amsthm,amstext,amscd,eucal,graphicx , xcolor}
\usepackage{slashbox}
\usepackage[all]{xy}
\usepackage{dsfont}
\usepackage{hyperref}
\usepackage{amsmath}
\usepackage{slashed}
\makeatletter \@addtoreset{equation}{section}

\makeatletter\renewcommand\section{\@startsection {section}{1}{\z@}%
                                   {-3.5ex \@plus -1ex \@minus -.2ex}
                                   {2.3ex \@plus.2ex}%
                                   {\normalfont\large\bfseries}}
\renewcommand\subsection{\@startsection{subsection}{2}{\z@}%
                                     {-3.25ex\@plus -1ex \@minus -.2ex}%
                                     {1.5ex \@plus .2ex}%
                                     {\normalfont\bfseries}}

\newcommand{\be}{\begin{equation}}
\newcommand{\ee}{\end{equation}}
\newcommand{\bea}{\begin{eqnarray}}
\newcommand{\eea}{\end{eqnarray}}
\newcommand{\bse}{\begin{subequations}}
\newcommand{\ese}{\end{subequations}}
\newcommand{\beqa}{\begin{eqnarray}}
\newcommand{\eeqa}{\end{eqnarray}}
\newcommand{\beqar}{\begin{eqnarray*}}
\newcommand{\eeqar}{\end{eqnarray*}}
\newcommand{\bi}{\begin{itemize}}
\newcommand{\ei}{\end{itemize}}
\newcommand{\bn}{\begin{enumerate}}
\newcommand{\en}{\end{enumerate}}

\newcommand{\ba}{\begin{array}}
\newcommand{\ea}{\end{array}}
\newcommand{\bc}{\begin{center}}
\newcommand{\ec}{\end{center}}

\def\bgs{Ba\~nados geometries}
\def\zhp{\zeta_{_{\mathcal{H}_+}}}
\def\zhm{\zeta_{_{\mathcal{H}_-}}}
\def\zhpm{\zeta_{_{\mathcal{H}_\pm}}}
\def\zhmp{\zeta_{_{\mathcal{H}_\mp}}}
\newcommand{\ro}[1]{r_{1{#1}}}
\newcommand{\rt}[1]{r_{2{#1}}}

\definecolor{darkgreen}{rgb}{0,0.3,0}
\definecolor{darkblue}{rgb}{0,0,0.3}
\definecolor{darkred}{rgb}{0.7,0,0}

\begin{document}

\begin{titlepage}

\begin{flushright}\vspace{-3cm}
{
\today }\end{flushright}
\vspace{-.5cm}

\begin{center}
\Large{\bf{On 3d Bulk Geometry of Virasoro Coadjoint Orbits:}}

\large{\bf{Orbit invariant charges and Virasoro hair on locally AdS$_3$ geometries}}

\bigskip\bigskip

\large{\bf{M.M. Sheikh-Jabbari\footnote{e-mail:
jabbari@theory.ipm.ac.ir}$^{;\ a}$
and  H. Yavartanoo\footnote{e-mail: yavar@itp.ac.cn }$^{;\ b}$  }}
\\

\vspace{5mm}
\normalsize
\bigskip\medskip
{$^a$ \it School of Physics, Institute for Research in Fundamental
Sciences (IPM),\\ P.O.Box 19395-5531, Tehran, Iran}
\smallskip

{$^b$ \it  State Key Laboratory of Theoretical Physics, Institute of Theoretical Physics,
Chinese Academy of Sciences, Beijing 100190, China.
 }\\

\end{center}
\setcounter{footnote}{0}

\begin{abstract}
\noindent
Expanding upon [arXiv:1404.4472, 1511.06079], we provide further detailed analysis of Ba\~nados geometries, the most general solutions to the AdS$_3$ Einstein gravity with Brown-Henneaux boundary conditions. We analyze in some detail the causal, horizon and boundary structure, and geodesic motion on these geometries, as well as the two class of symplectic charges one can associate with these geometries: charges associated with the exact symmetries and the Virasoro charges.
We elaborate further the one-to-one relation between the coadjoint orbits of two copies of Virasoro group  and Ba\~nados geometries. We discuss that the information about the Ba\~nados goemetries fall into two categories: ``orbit invariant'' information and ``Virasoro hairs''. The former are geometric quantities  while the latter are specified by the non-local surface integrals. We  elaborate on multi-BTZ geometries which have some number of disconnected pieces at the horizon bifurcation curve. We study multi-BTZ black hole thermodynamics and discuss that the thermodynamic quantities are orbit invariants. We also comment on the implications of our analysis for a 2d CFT dual which could possibly be dual to AdS$_3$ Einstein gravity.
\end{abstract}

\end{titlepage}
\renewcommand{\baselinestretch}{1.1}  
\tableofcontents

\section{Introduction}

Three dimensional gravity has very interesting features and has been long viewed as a testing ground for ideas in semiclassical and quantum gravity \cite{3d-gravity}. 3d Einstein gravity on flat (or AdS$_3$, {dS$_3$}) backgrounds do not have propagating degrees of freedom and all solutions to these theories are locally flat (or AdS$_3$, {dS$_3$}). Nevertheless, it is well known that they admit nontrivial solutions \cite{3d-solns}, including black holes \cite{BTZ}. 
One can in fact classify all the solutions to these theories with  prescribed boundary conditions. In this paper we will focus on the AdS$_3$ gravity case and analyze  family of locally AdS$_3$ geometries with Brown-Henneaux  boundary conditions \cite{Brown-Henneaux}. 

These solutions, the  Ba\~nados geometries, which were first discussed in \cite{Banados}, are specified by two arbitrary periodic functions $L_{+}=L_+(x^+)$  and $L_{-}=L_-(x^-)$, $L_\pm(x^\pm+2\pi)=L_\pm(x^\pm)$. 
Although all Ba\~nados geometries are locally AdS$_3$ and locally diffeomorphic to each others, it is now established that these geometries are physically distinct, as one can specify them with quasi-local, conserved surface charges, see \cite{AdS3-charges-everywhere} and references therein for a recent detailed study. This latter is also reflected in the fact that there are  no everywhere smooth coordinate transformations which respect the periodicity in $x^\pm$ and can bring theses geometries to global or Poincar\'e patch AdS$_3$ \cite{AdS3-charges-everywhere} (see also \cite{Rooman, Garbarz}).

In fact, one can distinguish two kinds of such conserved charges: those associated with \emph{exact symmetries} (Killing symmetries) of the Ba\~nados solutions and those which are in the family of \emph{symplectic symmetries}.\footnote{Note that the notion of symplectic symmetries extend the notion of asymptotic symmetries discussed in \cite{Brown-Henneaux}, in the sense that the charges are not only defined at the AdS$_3$ boundary, but also on any codimension two, compact, spacelike curve in the bulk.}  See \cite{AdS3-charges-everywhere, Hajian:2015xlp} for a detailed discussion on the concepts and the terminology. 

If we denote the generators of the exact symmetries by $J_\pm$ and those of symplectic symmetries by $L_n, \bar L_n$, one can show that the latter form two (left and right) copies of Virasoro algebras at Brown-Henneaux central charge $c$ \cite{Brown-Henneaux} and that $J_\pm$ commute with each other and also with all $L_n, \bar L_n$ \cite{AdS3-charges-everywhere}. Since these two sets are commuting, one may hence label the geometries by  $J_\pm$ as well as the Virasoro charges. As discussed in \cite{AdS3-charges-everywhere} (see also \cite{Garbarz, Mitchell}) one may then classify Ba\~nados geometries by the product of the two, left and right, Virasoro coadjoint orbits \cite{Witten-88, Balog}. Each Virasoro coadjoint orbit which is in correspondence with a {``representation'' (an orbit)} of the Virasoro group, is generically labelled by an integer and a continuous real number\footnote{As we will discuss in detail in section \ref{sec-4}, these two labels are not enough to uniquely specify the orbit and we need to also specify the type of the orbit.} and then ``states'' in a given orbit are fully specified once we also give their Virasoro charges, the ``Virasoro hairs''. The goal of this paper is to elaborate further on the results of \cite{AdS3-charges-everywhere, Banados-Killings} and on the one-to-one relation between two copies of Virasoro orbits and Ba\~nados geometries. 

The picture we depict here will correct and complete the one given in \cite{Banados-Killings}: The information about Ba\~nados geometries available to local observables of the usual classical GR, the geometric notions such as geodesic length, causal and boundary structure and the black hole (thermo)dynamics quantities like surface gravity and horizon angular velocity, entropy,  are ``orbit invariant quantities''. That is, all geometries which fall into the same orbit share these properties, regardless of their ``Virasoro hair''. On the other hand, the information about the Virasoro hair are semiclassical, in the sense that they are of the form of surface non-local (``quasi-local'') charges; the Virasoro charges may be viewed as the ``hair'' on classical geometries which all have the same mass, angular momentum and causal structure. Given this picture, one may then hope to obtain a full quantum description upon quantization of Virasoro coadjoint orbits. We shall provide some discussions on the latter point in the end, a thorough analysis is left to upcoming works.

To this end, in section \ref{sec-2} we analyze geometric aspects of Ba\~nados solutions. This includes reviewing their Killing vectors \cite{Banados-Killings}, analyzing the horizon, {causal}  and boundary structure of these geometries and geodesic motion on these geometries. In section \ref{sec-3}, we analyze charges associated with Ba\~nados metrics. In section \ref{sec-4}, {after reviewing} Virasoro coadjoint orbits and their classification, we discuss the 3d Ba\~nados geometry associated with Virasoro coadjoint orbits. The last section is devoted to a summary and outlook.
In an appendix we have given {detailed analysis} of special, but important cases, the geometries corresponding to Virasoro coadjoint orbits with constant character representative.

\section{Ba\~nados geometries and their causal structure}\label{sec-2}

In this paper we focus on the most general solutions to the AdS$_3$ Einstein gravity equations, 
\be\label{3d-Einstein}
R_{\mu\nu}=-\frac{2}{\ell^2} g_{\mu\nu}\,,
\ee
with Brown-Henneaux boundary conditions \cite{Brown-Henneaux}. These solutions are all locally AdS$_3$ with local $sl(2,\mathbb{R})\times sl(2,\mathbb{R})$ isometries. The Ba\~nados solutions in the Fefferman-Graham gauge \cite{FG-gauge} is given by \cite{Banados}
\be\label{generic-solutions}
ds^2=\ell^2\frac{ dr^2}{r^2}-(rdx^+-\frac{\ell^2}{r}L_- dx^-)(rdx^--\frac{\ell^2}{r}L_+dx^+)\,,
\ee
where $L_{+}=L_+(x^+)$  and $L_{-}=L_-(x^-)$ are arbitrary smooth, periodic functions, $L_\pm(x^\pm+2\pi)=L_\pm(x^\pm)$. We assume here that $x^\pm\in [0,2\pi]$ and parametrize two circles. The $x^\pm,\ r$ coordinate system used in \eqref{generic-solutions} will be called Ba\~nados  coordinate system, or Ba\~nados gauge. Similar solution in other gauges, e.g. in the Gaussian null coordinates (also called BMS gauge) has been constructed and analyzed \cite{ AdS3-charges-everywhere,soln-GNC-gauge}. For metrics \eqref{generic-solutions}, 
\be\label{det-g}
\det g=-\frac{\ell^2}{4r^6}(r^4-r_0^4)^2,\qquad r_0^4\equiv \ell^4L_+L_-.
\ee

{\paragraph{On the range of $r$ coordinate.} As we see the metric \eqref{generic-solutions} may be written in terms of $r^2$ (and not just $r$). Therefore, in principle, $r^2$ can assume positive or negative real values.} However, at large {$|r^2|$}, the metrics \eqref{generic-solutions} take the form 
$$
ds^2\approx  \ell^2\frac{ dr^2}{r^2}-r^2dx^+ dx^-,
$$
and without loss of generality one may choose $x^\pm=\tau\pm\varphi$, where $\varphi\in [0,2\pi]$ is a spacelike circle while $\tau$ is a timelike coordinate. This is in accord with the fact that causal boundary of the geometry is (part of) the cylinder parametrized by $x^\pm$.  With this choice, to avoid appearance of Closed Timelike Curves (CTC),  {$|r^2|$ cannot take a large negative value. We shall cut the $r^2$ range from a negative value, $r^2_{CTC1}$, where CTC develops and take $r^2_{CTC1}<r^2$. This range of $r^2$, however, is not necessarily CTC free. As we will argue in the next section, there is still a range $r^2_{CTC2}<r^2<0$, with $r^2_{CTC1}\leq r^2_{CTC2}<0$, where we have CTC. We should cut this range out too. The acceptable range of $r^2$ will then contain two disconnected pieces: $r^2_{CTC1}\leq r^2\leq r^2_{CTC2}<0$ and $r^2>0$. }
Moreover, as we will argue this allowed range of $r^2$ covers the geometry twice.  As a rough argument for the latter, note that if $r_0^4>0$ then in $r^2\ll 1$ region the metric essentially takes the same form as $r^2\gg 1$ region. {This will become more clear in the level Penrose diagrams shown in the next section. For illustrative purposes we have discussed the special case of BTZ black hole \cite{BTZ}, corresponding to constant positive $L_\pm$ in the appendix and compared the relation between Ba\~nados and the more standard BTZ-coordinate systems.} 

\subsection{Diffeomorphisms preserving the Ba\~nados gauge}\label{sec-2-1}

The geometries \eqref{generic-solutions} are written in a specific coordinate system, the Ba\~nados gauge. This coordinate system, extends the Fefferman-Graham coordinates (which are usually defined near the boundary at large $r$),  to arbitrary $r$. The Ba\~nados gauge is hence defined by the ``gauge fixing conditions''
\be\label{Banados-gauge}
g_{rr}=\frac{1}{r^2},\qquad  g_{r+}=g_{r-}=0.
\ee
One may then ask what is the ``residual'' diffeomorphisms \cite{Residual} which preserve the Ba\~nados gauge. This question was explored and answered in \cite{AdS3-charges-everywhere}:
\begin{subequations}\label{BG-preserving-diff}
\begin{align}
\chi[\epsilon_+,\epsilon_-]&=\chi^r \partial_r+\chi^+\partial_++\chi^-\partial_- ,\\
\chi^r=-\frac{r}{2} (\epsilon_+'+\epsilon_-'  ),\quad
\chi^+=\epsilon_+ +& \frac{\ell^2r^2\epsilon_-''+\ell^4L_-\epsilon_+''}{2(r^4-\ell^4L_+L_-)},\quad \chi^-=\epsilon_- + \frac{\ell^2r^2\epsilon_+''+\ell^4L_+\epsilon_-''}{2(r^4-\ell^4L_+L_-)},
\end{align}
\end{subequations}
where $\epsilon_{\pm}=\epsilon_{\pm}(x^{\pm})$ are two \emph{arbitrary, periodic} functions, i.e. $\epsilon_{\pm}(x^{\pm})=\epsilon_{\pm}(x^{\pm}+2\pi)$ and  $prime$  denotes derivative with respect to the argument.

Although  $\chi$  diffeomorphisms keep the form of metric invariant, they generically shift functions $L_\pm$. From Lie-derivative of metric ${\cal L}_\chi g_{\mu\nu}$ one can read that \cite{AdS3-charges-everywhere}
\be\label{delta-L}
\delta_\chi L_+=-\frac12\epsilon_+^{'''} +\epsilon_+L_+' + 2\epsilon_+'L_+ ,\qquad \delta_\chi L_-=-\frac12\epsilon_-^{'''} +\epsilon_-L_-' +2 \epsilon_-'L_-.
\ee

\subsection{Killing vectors of Ba\~nados geometries}\label{sec-2-2} 

 Killing vectors $\zeta$, with ${\cal L}_{\zeta} g=0$, should have the form \eqref{BG-preserving-diff} but for specific $\epsilon$'s satisfying $\delta_\zeta L_\pm=0$. We will denote the corresponding $\epsilon$'s by $K_\pm$. Therefore, Killing vectors are of the form \cite{Banados-Killings}
\begin{subequations}\label{KV}
\begin{align}
\zeta[K_+,K_-]&=\zeta^r \partial_r+\zeta^+\partial_++\zeta^-\partial_-,\\
\zeta^r=-\frac{r}{2} (K_+'+K_-'  ),\quad
\zeta^+=K_+ +& \frac{\ell^2r^2K_-''+\ell^4L_-K_+''}{2(r^4-\ell^4L_+L_-)},\quad \zeta^-=K_- + \frac{\ell^2r^2K_+''+\ell^4L_+K_-''}{2(r^4-\ell^4L_+L_-)},
\end{align}
\end{subequations}
where now $K_{\pm}$ should satisfy following equations
\be\label{KV-conditions}
\begin{split}
K_+^{'''} - 4 K_+'L_+ - 2K_+L_+'&=0,\cr
K_-^{'''} - 4 K_-'L_- - 2K_-L_-'&=0\,.
\end{split}
\ee
Since the notion of Killing vector is a local one, every  solutions to \eqref{KV-conditions} generate a Killing vector, regardless of the fact that the corresponding $K_\pm$ are periodic or not.  The above third order equations have six linearly independent solutions. These solutions constitute the six local isometries of \bgs\ which satisfy $sl(2,\mathbb{R})\times sl(2,\mathbb{R})$ algebra, as expected, recalling that Ba\~nados geometries are locally AdS$_3$ \cite{Banados-Killings}. $K_\pm(x^\pm)$ functions which solve \eqref{KV-conditions} are not necessarily periodic. Therefore, not all of the associated six Killing vector field are ``globally defined''. 

It was noted in \cite{Balog, Banados-Killings} that \eqref{KV-conditions} may be solved through a second order Schr\"odinger type equation:
\be\label{Schrodinger}
\psi''-L_+\ \psi=0\,,\qquad \phi''-L_-\ \phi=0.
\ee
If the two linearly independent solutions to the above are denoted as $\psi_1,\ \psi_2$ and $\phi_1,\phi_2$, then it is easy to verify that
\be\label{Ka}
\begin{split}
K_{+}^-\equiv \frac12\psi_1^2\,,\qquad K_{+}^+\equiv \frac12\psi_2^2\,,\qquad
K_{+}^0\equiv \frac12\psi_1\psi_2\,,\cr
K_{-}^+\equiv \frac12\phi_1^2\,,\qquad K_{-}^+\equiv \frac12\phi_2^2\,,\qquad
K_{-}^0\equiv \frac12\phi_1\phi_2\,,
\end{split}\nonumber
\ee
provide the three linearly independent solutions to \eqref{KV-conditions} where we adopt the normalization 
\be\label{normalization}
\psi'_1\psi_2-\psi_1\psi'_2=1\,,\qquad \phi'_1\phi_2-\phi_1\phi'_2=1. 
\ee  

Recalling that $L_\pm$ are periodic smooth functions, Floquet theorem implies that \cite{Banados-Killings, Hill-Eq} 
\be\label{psi-phi-Floquet}
\begin{split}
\psi_1= e^{{\mathcal T}_+ x^+} P_1(x^+)\,,\qquad \psi_2= e^{-{\mathcal T}_+ x^+} P_2(x^+)\,,\\
\phi_1= e^{{\mathcal T}_- x^-} Q_1(x^-)\,,\qquad \phi_2= e^{-{\mathcal T}_- x^-} Q_2(x^-)\,,
\end{split}\ee
where $P_i, Q_i$ are periodic smooth functions  and ${\mathcal T}_\pm$ are two constants built from $L_\pm$. In general ${\mathcal T}_\pm$ are two complex numbers and without loss of generality we can take $Re({\mathcal T}_\pm)\geq 0$. Discussions of \cite{Hill-Eq} reveals that, although $P_i, Q_i$ are generically $4\pi$ periodic, $P_1P_2$ and $Q_1Q_2$, and hence combinations like $\psi_1\psi_2, \psi_1/\psi_2, \psi_1'/\psi_1$ and $\psi_2'/\psi_2$ which have geometric meaning, are all $2\pi$ periodic. Therefore, in general case with $Re({\mathcal T}_\pm)\neq 0$, only two of the six Killings, those generated by $K_+^0,\ K_-^0$, are periodic. Hereafter, we will use the notation
\be\label{Kpm}
K_+=\psi_1\psi_2\,,\qquad K_-=\phi_1\phi_2\,,
\ee
for the functions generating the two periodic global Killing vectors which will be denoted by $\zeta_+$ and $\zeta_-$. In other words, the \bgs\ in general have $U(1)_+\times U(1)_-$ compact isometries.\footnote{There are  special $Re({\mathcal T}_+)$  and/or $Re({\mathcal T}_-)=0$ cases where more Killings are periodic. These cases may have four or six global Killings. The latter happens only for global AdS$_3$ with $L_\pm=-1/4$,  ${\mathcal T}_\pm=0$ where we have $Sl(2,\mathbb{R})\times Sl(2,\mathbb{R})$ global isometry.} 

The above construction makes it clear that solutions to \eqref{Schrodinger}, and hence solutions to \eqref{KV-conditions}, are implicit functions of $L_\pm$, i.e. $K_+=K_+(L_+(x^+)),\ K_-=K_-(L_-(x^-))$. In particular, we note that $K_\pm$ are only functions of $L_\pm$ and not their derivatives $L_\pm'$.

\subsection{Horizon and asymptotic boundary behavior}\label{sec-2-3}

Any linear combination of the two global Killing vector fields $\zeta_\pm$ is also a Killing vector. One may easily check that with the normalization \eqref{normalization} and \eqref{Kpm}, $|\zeta_+|^2=|\zeta_-|^2=\ell^2/4$, while $\zeta_+\cdot \zeta_-\neq 0$ is a nontrivial function. There is a certain combination of the two Killings which are normal to each other:
\be\label{zeta-Hs}
\zeta_{_{\mathcal{H}_\pm}} \equiv  \zeta_+ \pm\zeta_-\;.
\ee
The norm of these vectors are given as
\be\label{norm-zeta-Hs}
|\zeta_{_{\mathcal{H}_\pm}}|^2=\mp\frac{\psi_1\psi_2\phi_1\phi_2}{r^2}(r^2-\ro{\pm}^2)(r^2-\rt{\pm}^2),
\ee
where 
\be\label{rHs}
\ro{+}^2=\ell^2\frac{\psi_1'\phi_1'}{\psi_1\phi_1},\quad \rt{+}^2=\ell^2\frac{\psi_2'\phi_2'}{\psi_2\phi_2},\quad 
\ro{-}^2=\ell^2\frac{\psi_1'\phi_2'}{\psi_1\phi_2},\quad\rt{-}^2=\ell^2\frac{\psi_2'\phi_1'}{\psi_2\phi_1}\;.
\ee
For later use we note that {$\ro{+}^2\rt{+}^2=\ro{-}^2\rt{-}^2$} and  
\be\label{rHs-useful-relations}
\begin{split}
\ro{+}^2-\ro{-}^2=\frac{\ell^2}{\phi_1\phi_2}\frac{\psi_1'}{\psi_1},&\qquad \rt{+}^2-\rt{-}^2=-\frac{\ell^2}{\phi_1\phi_2}\frac{\psi_2'}{\psi_2},\\
\ro{+}^2-\rt{-}^2=\frac{\ell^2}{\psi_1\psi_2}\frac{\phi_1'}{\phi_1},&\qquad\rt{+}^2-\ro{-}^2=-\frac{\ell^2}{\psi_1\psi_2}\frac{\phi_2'}{\phi_2}\;.
\end{split}
\ee
Using these identities one may readily check that 
\be\label{zetaPhi-zetaH}
\zeta_{_{\mathcal{H}_+}}\cdot \zeta_{_{\mathcal{H}_-}}=0\;,\qquad |\zeta_{_{\mathcal{H}_\pm}}|^2\bigg|_{|\zeta_{_{\mathcal{H}_\mp}}|^2=0}=\ell^2\;.
\ee

\paragraph{Note:} Although most of the statements in this subsection is also true for generic ${\mathcal T}_\pm$ (\emph{cf.} \eqref{psi-phi-Floquet}), in sections \ref{sec-2} and \ref{sec-3}, we will be assuming that ${\mathcal T}_\pm$  are real-valued.

\subsubsection{Killing horizons and bifurcation surfaces}\label{sec-2-3-1}  

At the surfaces where the Killing vectors $\zhpm$ become null we have Killing horizons. Explicitly, either of the four codimension one surfaces $\mathcal{H}_{\alpha \sigma}$,
\be
\mathcal{H}_{\alpha \sigma}:\quad r^2=r^2_{\alpha\sigma},\quad \alpha=1,2, \ \sigma=+, -, 
\ee
define a null surface along the corresponding null Killing vector. 

We will show below that 
{
\begin{itemize}
\item if $r^2_{\alpha\sigma}$ is in the acceptable range (where geometry is CTC-free, \emph{cf.} discussion in the paragraph below \eqref{det-g}), then we have Killing horizons.
\item  These horizons are generically (not always)  \emph{bifurcate}-Killing horizons. We specify what is the bifurcation curve.
\item When they exist, these Killing horizons are either event horizon (outer horizon) or Cauchy horizon (inner horizon); if the event horizon is generated by $\zeta_{_{\mathcal{H}_\pm}}$,  the inner horizon is generated by $\zeta_{_{\mathcal{H}_\mp}}$. 
\item Then ${\cal H}_{\alpha\pm}$ null surfaces for $\alpha=1,2$ correspond to two branches of the event horizon and vice-versa, and the inner and outer bifurcation curves are  ${\cal H}_{1\pm}\cap {\cal H}_{2\pm}$, and are given by the equation $r^2=r^2_{1\pm}=r^2_{2\pm}$. 
\end{itemize}
}

To ensure existence of horizons and whether they are inner or outer horizon, we need to analyze the signs, zeros and infinities of $r^2_{\alpha\pm}$ and $|\zeta_{_{\mathcal{H}_\pm}}|^2$. {As discussed, range of $r$ coordinate can contain $r^2<0$ regions. The CTC appears in the regions where both of  $\zeta_{_{\mathcal{H}_\pm}}$ are spacelike. If the condition for existence of horizons cannot be met in all of the $x^\pm\in[0,2\pi]$ region, we are forced to cut some parts (due to CTC).} Moreover, since in general $\psi$ and $\phi$ have zeros, the horizon and boundary regions are not simply connected and will generically have some disconnected patches. Another point we will discuss below is that the allowed range for $r^2$ gives a double cover of spacetime. This latter, the fact that the $r^2<0$ should be also included in range of coordinate and the existence of CTC's  may be seen very explicitly for the BTZ case of constant, positive $L_\pm$ which is discussed in the Appendix A.1. The analysis of causal structure will be quite different for constant or nonconstant cases and they need to be discussed separately. 


\paragraph{Case I: constant $L_\pm$.} This case is the more studied and better understood case, e.g. see \cite{Banados-Killings, AdS3-charges-everywhere} and references therein. For the case where $L_\pm\geq 0$ we have BTZ black holes. This case will be discussed in some detail in the Appendix. For the BTZ case, as is well-known \cite{BTZ}, we  generically have a (bifurcate) simply connected horizon which is a circle.

\paragraph{Case II: generic $L_\pm$ case.} Unlike Case I, in this case $\psi$ and $\phi$ functions can have roots and zeros. Here and below unless mentioned explicitly, we will be considering nonconstant generic $L_\pm$ cases. Before starting  the analysis, we note the facts that, as implied by \eqref{Schrodinger} and \eqref{normalization}, 
\bi\item $\psi_1, \psi_2$ are smooth functions and can only have  simple roots.
\item  $\psi'_1, \psi'_2$ are also smooth and can only have simple roots.
\item Number of simple roots of $\psi_1$ in $x^+\in[0,2\pi]$ is equal to the number of roots of $\psi_2$ in the same range. Let us denote this number by $n_+$.
One may show that $\psi'_1,\psi'_2$ have the same number of roots \cite{Hill-Eq}. 
\item Between any two root of $\psi_1$ ($\psi_2$), there is a root of $\psi_2$ ($\psi_1$), and similarly for their derivatives \cite{Hill-Eq}.
\item If roots of $\psi_2,\psi'_1,\psi'_2,\psi_1$ are respectively denoted by $x^+_{1,i},x^+_{2,i},x^+_{3,i},x^+_{4,i}$, $i=1,\cdots, n_+$, we have
$x^+_{1,i}<x^+_{2,i}<x^+_{3,i}<x^+_{4,i}$ and that $x^+_{4,i}<x^+_{1,i+1}$. Note that we are using normalization \eqref{normalization}.
\item Considering the roots, we can divide the $[0,2\pi]$ range into $4n_+$ regions:
\be
\begin{split}
I_{1,1}&=[0,x^+_{1,1}),\ I_{2,1}=(x^+_{1,1},x^+_{2,1}),\ I_{3,1}=(x^+_{2,1},x^+_{3,1}),\ I_{4,1}=(x^+_{3,1},x^+_{4,1}),\\ 
I_{1,2}&=(x^+_{4,1},x^+_{1,2}),\ I_{2,2}=(x^+_{1,2},x^+_{2,2}),\cdots, I_{3,n_+}=(x^+_{2,n_+},x^+_{3,n_+}),\ I_{4,n_+}=(x^+_{3,n_+},x^+_{4,n_+}). 
\end{split}
\ee
The range $(x^+_{4,n_+},2\pi]$ is identified with the $I_{1,1}$ region.

\item Let us focus on the $i^{th}$ roots, i.e. the $I_{a,i}, a=1,2,3,4$ regions. One can always choose the overall sign of $\psi_1$ and $\psi_2$ functions such that
\begin{center}
\begin{tabular}{|c|c|c|c|c|c|}
\hline  \backslashbox{\small{function}}{\small{region}}&  $I_{1,i}$& $I_{2,i}$ &  $I_{3,i}$ & $I_{4,i}$ & $I_{1,i+1}$ \\
\hline
$\psi_1$ &  $+$ & $+$ & $+$ &$+$ & $-$ \\
$\psi_2$ &  $+$ & $-$ & $-$ & $-$ & $-$ \\
$\psi'_1$ & $+$  & $+$   & $-$ & $-$ & $- $\\
$\psi'_2$ & $-$ & $-$ & $-$ & $+$ & $+$ \\
\hline
$\psi_1\psi_2$ & $+$ & $-$ & $-$ &$-$ &$+$ \\
$\psi'_1/\psi_1$ & $+$& $+$& $-$ &$-$ &$+$ \\
$\psi'_2/\psi_2$ &$-$ &$+$ &$+$ &$-$ &$-$ \\
\hline
\end{tabular}
\end{center}

Note that for $2\pi$ periodic functions like $\psi_1\psi_2$ and $\psi'_\alpha/\psi_\alpha$, the first and fifth column are the same.
 \item In a similar fashion $\phi$ functions may have $n_-$ roots, with the same properties and ordering.
\ei 

\paragraph{On the existence of horizons.} The horizons, if exist, should be at the roots of $|\zhpm|^2$. To distinguish which one is the inner horizon and which one the outer, we need to study the sign of $r^2_{\alpha\pm}$ functions. 
Given the above analysis on the roots and signs of $\psi,\phi$'s, we learn that $r_{1\pm}^2, r_{2\pm}^2$ in the $I^+_{a,i}$ and $I^-_{b,j}$ regions, have the signs given in tables \ref{Table-1}.
\begin{table}[h]
\begin{center}
\begin{tabular}{|c|c|c|c|c|c|}
\hline \backslashbox{$x^+$}{$x^-$}&  $I^-_{1,j}$& $I^-_{2,j}$ &  $I^-_{3,j}$ & $I^-_{4,j}$ & $I^-_{1,j+1}$ \\
\hline
$I^+_{1,i}$ & $+$ & $+$ & $-$ & $-$ & $+$ \\
\hline
$I^+_{2,i}$ & $+$ & $+$ & $-$ & $-$ & $+$ \\
\hline
$I^+_{3,i}$ & $-$  & $-$   & $+$ & $+$ & $-$\\
\hline
$I^+_{4,i}$ & $-$  & $-$   & $+$ & $+$ & $-$ \\
\hline
$I^+_{1,i+1}$ & $+$ & $+$ & $-$ & $-$ & $+$ \\
\hline
\end{tabular}
\begin{tabular}{|c|c|c|c|c|c|}
\hline \backslashbox{$x^+$}{$x^-$}&  $I^-_{1,j}$& $I^-_{2,j}$ &  $I^-_{3,j}$ & $I^-_{4,j}$ & $I^-_{1,j+1}$ \\
\hline
$I^+_{1,i}$ & $+$ & $-$ & $-$ & $+$ & $+$ \\
\hline
$I^+_{2,i}$ & $-$ & $+$ & $+$ & $-$ & $-$ \\
\hline
$I^+_{3,i}$ & $-$ & $+$ & $+$ & $-$ & $-$ \\
\hline
$I^+_{4,i}$ & $+$ & $-$ & $-$ & $+$ & $+$ \\
\hline
$I^+_{1,i+1}$ & $+$ & $-$ & $-$ & $+$ & $+$\\
\hline
\end{tabular}\\
\begin{tabular}{|c|c|c|c|c|c|}
\hline \backslashbox{$x^+$}{$x^-$}&  $I^-_{1,j}$& $I^-_{2,j}$ &  $I^-_{3,j}$ & $I^-_{4,j}$ & $I^-_{1,j+1}$ \\
\hline
$I^+_{1,i}$ & $-$ & $+$ & $+$ & $-$ & $-$ \\
\hline
$I^+_{2,i}$ & $-$ & $+$ & $+$ & $-$ & $-$ \\
\hline
$I^+_{3,i}$ & $+$  & $-$   & $-$ & $+$ & $+$\\
\hline
$I^+_{4,i}$ & $+$  & $-$   & $-$ & $+$ & $+$ \\
\hline
$I^+_{1,i+1}$ & $-$ & $+$ & $+$ & $-$ & $-$\\
\hline
\end{tabular}
\begin{tabular}{|c|c|c|c|c|c|}
\hline \backslashbox{$x^+$}{$x^-$}&  $I^-_{1,j}$& $I^-_{2,j}$ &  $I^-_{3,j}$ & $I^-_{4,j}$ & $I^-_{1,j+1}$ \\
\hline
$I^+_{1,i}$ & $-$ & $-$ & $+$ & $+$ & $-$ \\
\hline
$I^+_{2,i}$ & $+$ & $+$ & $-$ & $-$ & $+$ \\
\hline
$I^+_{3,i}$ & $+$ & $+$ & $-$ & $-$ & $+$ \\
\hline
$I^+_{4,i}$ & $-$ & $-$ & $+$ & $+$ & $-$ \\
\hline
$I^+_{1,i+1}$ & $-$ & $-$ & $+$ & $+$ & $-$\\
\hline
\end{tabular}
\caption{Left and Right top tables, respectively show sign of $r_{1+}^2$ and $r_{2+}^2$. Left and Right bottom tables respectively show sign of $r_{1-}^2$ and $r_{2-}^2$. Note that the change of sign for either of these functions happens at places where they vanish, or when they become infinite. The former happens at zeros of derivatives $\psi',\phi'$ while the latter at zeros of $\psi$ and $\phi$.}\label{Table-1}
\end{center}
\end{table}

\paragraph{The event horizon.} The outer (event) horizon is by definition the null surface which is the boundary of all the past or future light-cones of points at the AdS$_3$ boundary. For the AdS$_3$ case,  the Killing vector field generating the event horizon remains time-like at the boundary\footnote{We note that this is a generic property of AdS$_3$ black holes and is unlike the asymptotic flat black holes like Kerr. In the Kerr case the horizon generating Killing vector is generically spacelike in the asymptotic region of the spacetime, and become timelike only in a region very close to the axis of rotation (usually denoted by $\theta=0,\pi$).} (while becoming null at the horizon). In order to distinguish which of $r_{\alpha +}$ or $r_{\alpha -}$ gives the outer (event) horizons we need to distinguish which of $\zhpm$ are time-like at the boundary (large $r^2$ region). From \eqref{norm-zeta-Hs} we learn that for large $r^2$
\be
|\zeta_{_{\cal{H}_\pm}}|^2\approx \mp r^2\Pi, \qquad \Pi\equiv \phi_1\phi_2\psi_1\psi_2,
\ee
and the sign of ${\Pi}$ in different regions is given in the table \ref{Table-2-Pi}.
\begin{table}[h]
\begin{center}
\begin{tabular}{|c|c|c|c|c|c|}
\hline \backslashbox{$x^+$}{$x^-$}&  $I^-_{1,j}$& $I^-_{2,j}$ &  $I^-_{3,j}$ & $I^-_{4,j}$ & $I^-_{1,j+1}$ \\
\hline
$I^+_{1,i}$ & $+$ & $-$ & $-$ & $-$ & $+$ \\
\hline
$I^+_{2,i}$ & $-$ & $+$ & $+$ & $+$ & $-$\\
\hline
$I^+_{3,i}$ & $-$ & $+$ & $+$ & $+$ & $-$ \\
\hline
$I^+_{4,i}$ & $-$ & $+$ & $+$ & $+$ & $-$ \\
\hline
$I^+_{1,i+1}$ & $+$ & $-$ & $-$ & $-$ & $+$\\
\hline
\end{tabular}
\caption{Sing of $\Pi$. Sign of norm of $|\zeta_{_{\cal{H}_\pm}}|^2$ near the boundary is $\mp\Pi$.}\label{Table-2-Pi}
\end{center}
\end{table}

\paragraph{Inner and outer horizons and horizon radii differences.} The criterion above will distinguish which of the $r^2_{\alpha +}$ or $r^2_{\alpha -}$ correspond to bifurcate event horizon(s). One may distinguish which is the inner (Cauchy) horizon noting the following: The outer horizon is by definition the one which is closer to the boundary than the inner horizon. That is, the outer horizon should happen at a larger radius than the inner horizon. The signs of horizon radii differences are given in table \ref{Table-3-r-differences}.
\begin{table}[h]
\begin{center}
\begin{tabular}{|c|c|c|c|c|c|}
\hline \backslashbox{$x^+$}{$x^-$}&  $I^-_{1,j}$& $I^-_{2,j}$ &  $I^-_{3,j}$ & $I^-_{4,j}$ & $I^-_{1,j+1}$ \\
\hline
$I^+_{1,i}$ & $+$ & $-$ & $-$ & $-$ & $+$ \\
\hline
$I^+_{2,i}$ & $+$ & $-$ & $-$ & $-$ & $+$ \\
\hline
$I^+_{3,i}$ & $-$  & $+$   & $+$ & $+$ & $-$ \\
\hline
$I^+_{4,i}$ & $-$  & $+$   & $+$ & $+$ & $-$ \\
\hline
$I^+_{1,i+1}$ & $+$ & $-$ & $-$ & $-$ & $+$\\
\hline
\end{tabular}
\begin{tabular}{|c|c|c|c|c|c|}
\hline \backslashbox{$x^+$}{$x^-$}&  $I^-_{1,j}$& $I^-_{2,j}$ &  $I^-_{3,j}$ & $I^-_{4,j}$ & $I^-_{1,j+1}$ \\
\hline
$I^+_{1,i}$ & $+$ & $+$ & $-$ & $-$ & $+$ \\
\hline
$I^+_{2,i}$ & $-$ & $-$ & $+$ & $+$ & $-$\\
\hline
$I^+_{3,i}$ & $-$ & $-$ & $+$ & $+$ & $-$ \\
\hline
$I^+_{4,i}$ & $-$ & $-$ & $+$ & $+$ & $-$ \\
\hline
$I^+_{1,i+1}$ & $+$ & $+$ & $-$ & $-$ & $+$\\
\hline
\end{tabular}\\
\begin{tabular}{|c|c|c|c|c|c|}
\hline \backslashbox{$x^+$}{$x^-$}&  $I^-_{1,j}$& $I^-_{2,j}$ &  $I^-_{3,j}$ & $I^-_{4,j}$ & $I^-_{1,j+1}$ \\
\hline
$I^+_{1,i}$ & $+$ & $-$ & $-$ & $+$ & $+$ \\
\hline
$I^+_{2,i}$ & $-$ & $+$ & $+$ & $-$ & $-$ \\
\hline
$I^+_{3,i}$ & $-$  & $+$   & $+$ & $-$ & $-$ \\
\hline
$I^+_{4,i}$ & $-$  & $+$   & $+$ & $-$ & $-$ \\
\hline
$I^+_{1,i+1}$ & $+$ & $-$ & $-$ & $+$ & $+$\\
\hline
\end{tabular}
\begin{tabular}{|c|c|c|c|c|c|}
\hline \backslashbox{$x^+$}{$x^-$}&  $I^-_{1,j}$& $I^-_{2,j}$ &  $I^-_{3,j}$ & $I^-_{4,j}$ & $I^-_{1,j+1}$ \\
\hline
$I^+_{1,i}$ & $+$ & $-$ & $-$ & $-$ & $+$ \\
\hline
$I^+_{2,i}$ & $-$ & $+$ & $+$ & $+$ & $-$\\
\hline
$I^+_{3,i}$ & $-$ & $+$ & $+$ & $+$ & $-$ \\
\hline
$I^+_{4,i}$ & $+$ & $-$ & $-$ & $-$ & $+$ \\
\hline
$I^+_{1,i+1}$ & $+$ & $-$ & $-$ & $-$ & $+$\\
\hline
\end{tabular}
\caption{Left and Right top tables, respectively show sign of $r_{1+}^2-r_{1-}^2$ and $r_{1+}^2-r_{2-}^2$. Left and Right bottom tables respectively show sign of $r_{2+}^2-r_{1-}^2$ and $r_{2+}^2-r_{2-}^2$. To deduce the above tables we have used \eqref{rHs-useful-relations}.}\label{Table-3-r-differences}
\end{center}
\end{table}

\paragraph{How to build full $4n_+\times 4n_-$ tables.} We note that, as discussed above, if in general functions $\psi$ and $\phi$ have respectively $n_+$ and $n_-$ zeros, then the above tables, instead of being $5\times 5$ should have $4n_+ \times 4n_-=16n_+ n_-$ regions. The way to build these bigger tables from those given here is simply copying the above tables $n_-$ times on the rows, identifying the last and first columns in each copy; and copying $n_+$ times the columns and then identifying the last and first rows in each copy.\footnote{As we can explicitly see, in our $5\times 5$ tables the first and fifth row and column are the same.} In the end, we should also identify the first row and the last row, and the first column and the last column. This last identification is to implement the $2\pi$ periodicity of $\Pi$ and $r^2_{\alpha\pm}$ functions. 

We also note that the tables here are denoting {cylinder spanned by $x^\pm$ with} $x^\pm\in[0,2\pi]$. These tables are hence, in fact, depictions of the AdS$_3$ boundary.

\subsubsection{More on causal and boundary structure}\label{sec-2-3-2} 

Equipped with the information of the signs of $|\zhpm|^2$ and $r^2_{\alpha\pm}$, we are now ready to build the full causal structure of the Ba\~nados geometries and discuss horizon properties like geometry of inner and bifurcation curves, and the horizon angular velocity and surface gravity. As we will discuss the geometry, the boundary and the event horizon in general consists of some number of causally disconnected pieces. 
{To gain a better intuition and picture, however, we would like to invite the reader to go through the appendix A.1 where we discuss the simpler case of constant $L_\pm$.}

\paragraph{Surface gravity.} 
 One may show
\be\label{zeta-H-kappa}
-\frac14|\nabla\zhpm|^2\bigg|_{|\zhpm|^2=0}=1,
\ee
implying that the ``un-normalized surface gravity'' at the Killing horizons are equal (up to a sign). We stress that this equation is true for all four choices of roots of $|\zhpm|$. To read the physical surface gravity and determining its sign, however, we need to fix the normalization of the Killing vectors.

\paragraph{Fixing the normalization of Killings.} It is well known that in order to read the horizon kinematical properties, like horizon angular velocity and surface gravity, one needs to choose an appropriate normalization for the corresponding Killing vectors. In particular, we are interested in finding the horizon properties associated with the outer horizon, the horizon causally connected to the AdS$_3$ boundary. The Killing vector generating this outer (event) horizon should remain time-like in the $r^2>r^2_{_{\cal{H}}}$ region. This outer horizon can be generated by 
$\zeta_{_{{\cal H}_+}}$ ($\zeta_{_{{\cal H}_{-}}}$) in the regions where $\psi_1\psi_2\phi_1\phi_2$ is negative (positive). If the outer horizon is generated by $\zeta_{_{{\cal H}_{+}}}$ ($\zeta_{_{{\cal H}_{-}}}$) the horizon radius is  $r^2_{\alpha +}$  ($r^2_{\alpha -}$), where $r^2=r^2_{1+}$ and $r^2=r^2_{2+}$ ($r^2=r^2_{1-}$ and $r^2=r^2_{2-}$) denote the two branches of the bifurcate horizon.  We will return to this point later.

To fix the normalization of horizon generating Killings, we focus on the regions where $\psi_1\psi_2$ and $\phi_1\phi_2$ are both positive, and hence the event horizon is generated by $\zeta_{_{{\cal H}_+}}$. Similar analysis may also be repeated for the other regions, including cases with negative $\Pi$, where the outer (event) horizon is generated by $\zeta_{_{{\cal H}_-}}$. Let us study the large $r$, asymptotic behavior of the Killing vector $\zeta_{_{{\cal H}_+}}$. One may readily see from \eqref{KV} that at large $r$
\be
\zeta_+\sim K_+ \partial_+\,,\qquad \zeta_-\sim K_- \partial_-\,.
\ee 
The ``appropriate'' normalization is hence the one in which $\zeta_\pm$ are along the coordinates.  Noting \eqref{normalization} and \eqref{Kpm} we learn that
\be
\zeta_+\sim \frac{1}{2{\mathcal T}_+}\partial_{X^+}\,,\qquad \zeta_-\sim \frac{1}{2{\mathcal T}_-}\partial_{X^-}\,,
\ee
where
\be\label{Xpm-b'dry-coord}
X^+=\frac{1}{2{\mathcal T}_+}\ln \frac{\psi_1}{\psi_2},\qquad X^-=\frac{1}{2{\mathcal T}_-}\ln \frac{\phi_1}{\phi_2}.
\ee
In the above the normalization factors $\frac{1}{2{\mathcal T}_\pm}$ are chosen recalling \eqref{psi-phi-Floquet}, such that $X^\pm$ are $2\pi$ periodic, explicitly, $X^\pm(x^\pm+2\pi)=X^\pm(x^\pm)+2\pi$. The appropriate asymptotic time and angular variable, $\tau,\varphi$ are hence
\be\label{asymptotic-time}
\tau=\ell(X^+ + X^-)/2,\qquad \varphi=(X^+ - X^-)/2.
\ee
With the above normalization and recalling \eqref{zeta-H-kappa}, we learn that physical surface gravity $\kappa$ is 
\be
\frac2\kappa=\frac{1}{{\mathcal T}_+}+\frac{1}{{\mathcal T}_-}.
\ee

\paragraph{${\cal H}_{1\pm}, {\cal H}_{2\pm}$ intersecting the boundary, number of disconnected pieces at the boundary.} 

Let us focus on the regions where $\psi_1\psi_2$ and $\phi_1\phi_2$ are both positive, and hence the event horizon is generated by $\zeta_{_{{\cal H}_+}}$. Similar analysis may also be repeated for the other regions. The ``horizon radii'' $r^2_{\alpha +}$ can range from minus infinity to plus infinity. The places $r^2_{1+}$ or $r^2_{2+}$ become infinite is where the horizon  ${\mathcal{H}}_{1+}$ or ${\mathcal{H}}_{2+}$ intersect the AdS$_3$ boundary. To be explicit, in our $5\times 5$ tables, the segments of the boundary are in $(1,1), (2,2)$ and in $(4,4), (5,5)$ parts of the table. From the tables we also see that there are $(3,3)$ and parts of $(2,2), (4,4)$ regions where $\zeta_{_{{\cal H}_+}}$ can have zeros at $r^2<r_0^2$ (\emph{cf}. \eqref{det-g}). 

In fact, one can readily see that extrema of functions $r^2_{\alpha\pm}$ happen at\footnote{{Note that at $\left(\frac{\psi'}{\psi}\right)'=0$,  $\left(\frac{\psi'}{\psi}\right)^2=L_+$, similarly for $\phi$ and hence the extremum of $\frac{\psi'}{\psi}\frac{\phi'}{\phi}$ occurs at $\sqrt{L_+L_-}$ .}}
\be\label{rH-extrema}
\partial_+ r^2_{\alpha\pm}=0,\ \partial_- r^2_{\alpha\pm}=0 \quad \Longrightarrow \quad r^2_{\alpha\pm}|_{extremum}=r_0^2.
\ee
The extrema is a minimum in regions where $r^2_{\alpha \sigma}$ can become very large (and become infinite at the boundaries of the region) and the maximum in the regions where at its boundaries $r^2_{\alpha \sigma}$ can become zero.
That is, in region $(3,3)$ we have a maximum; in regions $(1,1), (5,5)$ we have a minimum and in regions $(2,2)$ and $(4,4)$ we have  a maximum and a minimum.
One can show (e.g. see the Appendix A for the case of BTZ black holes) that $r^2>r_0^2$ and $0<r^2<r_0^2$ regions are geometrically the same, {corresponding to the regions I and I' on the Penrose diagram in Figure \ref{rho-vs-r-BTZ}.} The Ba\~nados coordinates cover this parts twice. So, we would not count the event horizons  appearing in the $0<r^2<r_0^2$ region as independent.

We next note that in the region $\Pi<0$, the outer horizon may be generated by $\zeta_{_{{\cal H}_-}}$. In these regions $r^2_{1 -}$ or $r^2_{2-}$ may become infinite. In terms of our $5\times 5$ tables, this happens at the boundaries of $(4,1)$ and $(5,2)$ regions (where $r^2_{1 -}$ becomes infinite) and at the boundaries of $(1,4)$ and $(2,5)$ (where $r^2_{2 -}$ becomes infinite). Regions $(2,3), (3,4)$ and $(3,2), (4,3)$ are again a repetition of these parts and we do not count them separately.

All in all, the above analysis shows that we have $(n_+ +1)(n_-+1)$  causally disconnected regions in the range $x^\pm\in[0,2\pi]$ at the boundary. These regions are separated by the roots of $\psi_1', \psi_2', \phi_1'$ and $\phi_2'$. {Note that this is exactly places where  $|\zeta_{_{{\cal H}_\pm}}|^2$ around small $r$ region changes sign. It is also instructive to recall the case of BTZ black hole and in particular Figure \ref{rho-vs-r-BTZ}.}

\paragraph{Regions enclosed by the boundary and the event horizon.}

Let us consider the case where the outer horizon is generated by $\zeta_{_{{\cal H}_+}}$, i.e. where $\Pi>0$. The two surfaces $r^2=r^2_{1+}$ and $r^2=r^2_{2+}$ are defining the two branches of the bifurcate (event) horizon.\footnote{In the BTZ case $r^2_{1+}=r^2_{2+}=r^2_+$, and as usual on the Penrose diagram (see Figure \ref{rho-vs-r-BTZ}) the event horizon is specified by $r=r_+$. Note, however, the difference between the BTZ and the Ba\~nados coordinate systems (\emph{cf.} the appendix A.1).}  As discussed there are also patches defined by the intersection of ${\cal H}_{1-}$ or ${\cal H}_{2-}$ and the boundary (in the $\Pi<0$ regions). There are $(n_++1)(n_-+1)$ causally disconnected regions bordered by the two branches of the event horizon and the boundary.

\paragraph{Bifurcation curve $\Sigma$.}

As in the previous discussion, let us focus on the regions where $\psi_1\psi_2$ and $\phi_1\phi_2$ are both positive, and hence the event horizon is generated by $\zeta_{_{{\cal H}_+}}$. Similar analysis may also be repeated for the other regions.  As mentioned the $\mathcal{H}_{1+}$ and $\mathcal{H}_{2+}$ are two dimensional null surfaces generated by the Killing vector field $\zeta_{_{\mathcal{H}_+}}$. In terminology of usual Penrose diagram or Kruskal coordinates, had we introduced $u,v$ null coordinates, that is,  $\mathcal{H}_{1+}$ and $\mathcal{H}_{2+}$ are along constant $u$ and constant $v$ surfaces. They can intersect on a spacelike one dimensional curve $\Sigma$ (which is at $u=v=0$). In the usual terminology $\Sigma_+$ is the bifurcation curve. 
Similarly to $\mathcal{H}_{1+}$ and $\mathcal{H}_{2+}$, $\Sigma_+$ is not necessarily simply connected and may have some disconnected pieces. Below we discuss some properties of the bifurcation curve $\Sigma_+$:
\bn
\item The Killing vector field which generates the horizon vanishes at $\Sigma_+$. In the region we are discussing the horizon is generated by $\zeta_{_{{\cal H}_+}}$ and 
\be
\zeta_{_{{\cal H}_+}}|_{_{\Sigma_+}}=0.
\ee
This may be checked by a straightforward computation at $\Sigma_+={\mathcal{H}}_{1+}\cap {\mathcal{H}_{2+}}$. To this end,  it is useful to note that at $\Sigma_+$:
\be\label{key-relation-at-Sigma}
\Sigma_+:\qquad {\psi_1'}{\psi_2}=-{\phi_2'}{\phi_1}\,,\quad {\psi_2'}{\psi_1}=-{\phi_1'}{\phi_2}\;.
\ee
In a similar way, one may check that the above is also true when the horizon is generated by $\zeta_{_{\mathcal{H}_-}}$ with the bifurcation curve $\Sigma_-$, $\Sigma_-={\mathcal{H}}_{1-}\cap {\mathcal{H}_{2-}}$. For the $\Sigma_-$, however, we have
$$
\Sigma_-:\qquad {\psi_1'}{\psi_2}={\phi_1'}{\phi_2}\,,\quad {\psi_2'}{\psi_1}={\phi_2'}{\phi_1}\;.
$$
\vskip 2mm
\item \textbf{$\Sigma_\pm$ is along the Killing vector $\zeta_{_{\mathcal{H}_\mp}}$}.
One can readily show that tangent to the curve at the intersection of  $r^2\equiv r_{\mathcal{H}_{1\pm}}^2$ and $r^2\equiv r_{\mathcal{H}_{2\pm}}^2$ surfaces is along $\zeta_{_{\mathcal{H}_\mp}}$. In other words, $\Sigma_\pm$ is generated by the flow of $\zeta_{_{\mathcal{H}_\mp}}$ at ${\cal H}_{1\pm}$ or ${\cal H}_{2\pm}$. Moreover, recalling \eqref{zetaPhi-zetaH}, we see that $\zeta_{_{\mathcal{H}_\mp}}$ are  spacelike at $\Sigma_\pm$.
\vskip 2mm

\item One may calculate the Ba\~nados metric \eqref{generic-solutions} at $\Sigma_+$, to obtain
\be\label{Sigma+-metric}
ds^2\big|_{\Sigma_+}=\ell^2 d\Phi_+^2\,, \qquad \Phi_+=\frac12 \ln\frac{\psi_1'\psi_1}{\phi_1'\phi_1}=-\frac12 \ln\frac{\psi_2'\psi_2}{\phi_2'\phi_2}\;.
\ee
To show the second equality above we have used \eqref{key-relation-at-Sigma}. The above, together with \eqref{zetaPhi-zetaH} implies that $\zeta_{_{\mathcal{H}_-}}|_{\Sigma_+}= \partial_{\Phi_+}$. 

If the outer horizon is generated by $\zeta_{_{\mathcal{H}_-}}$, with the bifurcation curve $\Sigma_-$, the metric at $\Sigma_-$ is given by
\be\label{Sigma--metric}
ds^2\big|_{\Sigma_-}=\ell^2 d\Phi_-^2\,, \qquad \Phi_-=\frac12 \ln\frac{\psi_1'\psi_1}{\phi_2'\phi_2}=-\frac12 \ln\frac{\psi_2'\psi_2}{\phi_1'\phi_1}\;.
\ee
The coordinate $\Phi_-$ is along the Killing vector $\zeta_{_{\mathcal{H}_+}}$.
\vskip 2mm

\item Using \eqref{psi-phi-Floquet} we learn that 
\be\label{Phi-coord}
\Phi_\pm={\mathcal T}_+x^+ \mp {\mathcal T}_- x^- + {\cal P}_\pm(x^+)+ {\cal Q}_\pm(x^-),
\ee
where ${\cal P}_\pm, {\cal Q}_\pm$ are periodic functions, ${\cal P}_\pm(x^+)= {\cal P}_\pm(x^++2\pi), {\cal Q}_\pm(x^-)= {\cal Q}_\pm(x^-+2\pi)$.  
\vskip 2mm

\item \textbf{Horizon angular velocity}: From \eqref{Phi-coord} and \eqref{Xpm-b'dry-coord} we learn that 
\be\label{Phi-explicit}
\Phi_\pm=R^\pm_{{\cal H}}(\varphi-\Omega^\pm_{\cal H} \frac{\tau}{\ell})+ ``periodic\ part^{,,},
\ee
where
\be\label{Omega-H}
R^\pm_{{\cal H}}={\mathcal T}_+ \pm{\mathcal T}_-,\qquad \Omega^+_{\cal H}=\frac{1}{\Omega^-_{\cal H}}=\frac{{\mathcal T}_+-{\mathcal T}_-}{{\mathcal T}_++{\mathcal T}_-}\;.
\ee
One may readily show that we have the same horizon angular velocity if we considered event horizon generated by $\zeta_{_{\mathcal{H}_-}}$ and the bifurcation curve $\Sigma_-$.\footnote{Note that if the event horizon is generated by $\zeta_{_{\mathcal{H}_-}}$, one should revisit the definition of $X^\pm$ \eqref{Xpm-b'dry-coord}, by taking either of $X^-$ or $X^+$ to minus themselves. In this way the coordinate covering the event horizon will always keep the form $\Phi_+$. As we will discuss, $\Phi_-$ will be the coordinate on the inner horizon.\label{footnote-6}}
\vskip 2mm
\item \textbf{Length of $\Sigma_+$, ``area'' of horizon.} As can be explicitly seen from \eqref{Sigma+-metric} and \eqref{Phi-explicit}, $\Sigma$ is a curve of finite length,
\be\label{horizon-area}
\mathcal{A}_{\cal H}= 2\pi \ell ({\mathcal T}_+ + {\mathcal T}_-),
\ee
One may readily see that $\Sigma_-$ has the same ``length'' as above.
\vskip 2mm
\item \textbf{Bifurcation curve $\Sigma$ has $(n_++1)(n_-+1)$ disconnected pieces.} The bifurcation curves $\Sigma_+$, as parametrized by $\Phi_+$ coordinate \eqref{Phi-coord}, consist of some disconnected pieces. To distinguish the disconnected pieces, we should check where the equations defining $\Sigma_+$ \eqref{key-relation-at-Sigma} have solutions and also where $\Phi$ is  real-valued. Given the information in our tables,  we learn that $\Sigma_+$ has the same number of disconnected pieces as the ${\cal H}_{\alpha +}$. Considering the outer (event) horizons generated by $\zeta_{_{\mathcal{H}_+}}$, the bifurcation curve of the event horizon has $(n_++1)(n_-+1)$ disconnected pieces. 

\en

\paragraph{Inner horizons.} Associated with any event horizon there is, generically, an inner horizon. The region ``inside'' the inner horizon region is bounded between the inner horizon (which is where we have a null Killing vector field) and the region we develop a CTC. Inside this region the horizon generating Killing vector should remain time-like. When the event horizon is generated by $\zhpm$ the inner (Cauchy) horizon is generated by $\zhmp$. We note that, as is seen in table \ref{Table-3-r-differences}, in the regions event horizon is given by $r^2_{2+}$ ($r^2_{1+}$), $r^2_{2+}-r^2_{2-}$ and  
$r^2_{2+}-r^2_{1-}$ ($r^2_{1+}-r^2_{1-}$ and $r^2_{1+}-r^2_{2-}$) are both positive, confirming the picture that outer horizons happen a larger $r$-coordinate value than the inner horizon. This statement is also true when the event (outer) horizons generated by $\zeta_{_{\mathcal{H}_-}}$.

As commented in footnote \ref{footnote-6}, regardless of whether the Killing horizon generating the inner horizon is $\zeta_{_{\mathcal{H}_\mp}}$, 
one may show that the coordinate along the inner horizon is $\Phi_-$ (\emph{cf.} \eqref{Sigma+-metric}, \eqref{Sigma--metric}).

\paragraph{Summary of subsection:} 
We argued that the two Killing vectors $\zhpm$ which are everywhere orthogonal, can become null on some different patches of spacetime. These null surfaces can intersect the boundary in different places. As discussed boundary and horizon, in general, have $(n_++1)(n_-+1)$ disconnected pieces. When the outer horizon is generated by $\zhpm$ the corresponding inner horizon is generated by $\zhmp$. Regardless of which $\zeta$ producing the horizon, the coordinate spanning the bifurcation curve of the event horizon is $\Phi_+$  and that of the inner horizon is $\Phi_-$. The length of the event horizon is always $2\pi\ell R^+_{\cal H}$ and that of inner horizon is $2\pi\ell R^-_{\cal H}$. {We have summarized the above information in the Penrose diagram \ref{Penrose-diagram}.}

\subsection{Geodesic motion on generic Ba\~nados geometries}\label{sec-2-4}

To gain a better intuition about the Ba\~nados geometries we present a brief analysis of geodesic motion on these backgrounds. Similar study has been performed for constant $L_\pm$ case, 
cases which include BTZ black holes, in the literature. For clarity and completeness we have presented those analysis in the appendix A.2. Here, we give a discussion for generic $L_\pm$ functions.

As discussed any Ba\~nados geometry generically has two global Killing vectors $\zeta_\pm$ and four local ones, altogether we have six Killing vectors which form $sl(2,\mathbb{R})\times sl(2,\mathbb{R})$ algebra. 
We may use these facts to construct the geodesics.  Let us consider the geodesic velocity vector $v^\mu=dx^\mu/ds$, where $s$ is the affine parameter on the geodesic. The velocity along Killing vectors is a constant of motion. Therefore, $P_{\pm}$,
\be
P_{\pm}=v\cdot \zhpm\,,
\ee
are constants of motion. Having two constants of motion, we can completely solve for the velocity vector using the fact that
\be
v^2=\sigma, \qquad \sigma=-1,0,+1\ \ \text{respectively for time-like, null and space-like geodesics.}
\ee
To write out the above equation explicitly, let us define an orthogonal basis, $\zhpm$ and $\eta$, with 
\be
\ell^{-1}\eta=\eta^r \partial_r+\eta^+\partial_++\eta^-\partial_-,
\ee
where, $\eta\cdot\zhpm=0$, and
\be\begin{split}
\eta^r =-\frac{K_+ K_-}{\ell^2 r} (r^4-r_{\mathcal{H}_1}^2r_{\mathcal{H}_2}^2)\,,\quad
{\eta^{\pm}}=K_\pm K'_\mp-\frac{\ell^2r^2{K_{\pm}'}({K_{\mp}''}-4L_{\mp}{K_{\mp}})+\ell^4{K_{\mp}'}({K_{\pm}''}-4L_{\pm}{K_{\pm}})}{2(r^4-l^4L_+L_-) },
\end{split}
\ee
and one may  show that
\be\label{eta-norm}
|\eta|^2=-|\zhp|^2|\zhm|^2.
\ee

We can then expand the velocity vector $v$ as
\be
v=\frac{P_+}{|\zhp|^2} \zhp +\frac{P_-}{|\zhm|^2}\zhm+ \frac{Z}{|\eta|^2}\eta,
\ee
where $Z=v\cdot \eta$ and 
\be
Z^2= |\eta|^2 (\sigma-\frac{P_+^2}{|\zhp|^2}-\frac{P_-^2}{|\zhm|^2})=|\eta|^2 \sigma+{P_+^2}{|\zhm|^2}+{P_-^2}{|\zhp|^2}.
\ee
What will be important in our further discussions in sections \ref{sec-4} and \ref{sec-5} is that $Z^2, P_\pm$  are diffeomorphism invariant quantities. Therefore, geodesic observers can only probe a part of the information encoded in the functions $L_\pm$ which specify the geometric properties of the background.

%
%

\subsection{Summary of Ba\~nados geometries and outlook of the section}\label{sec-2-5}

We argued that in general, irrespective of the details of functions $L_\pm$, Ba\~nados metrics \eqref{generic-solutions} have $U(1)_+\times U(1)_-$ global isometries. The corresponding Killing vectors may be constructed through the solutions to the Hill's equations \eqref{Schrodinger}, $\psi$ and $\phi$ functions. In general, for nonconstant $L_\pm$ cases, norm of Killing vector fields $\zhpm$ can vanish at four surfaces given in \eqref{rHs}. As discussed two of these four surfaces provide outer (event) bifurcate horizon while the other two, the bifurcate inner Killing horizon. The   Ba\~nados coordinate system, once the range $r^2<0$ is also included, covers the regions outside the outer horizon (all the way to the boundary) and the inside the inner horizon (all the way to the CTC region). It does not necessarily cover the (whole) region between the two horizons. Moreover, as discussed, in the allowed region for $r^2$, it provides a double cover of the part of spacetime in covers.\footnote{{It is instructive to see analysis of appendix A.1 for the simpler case of constant $L_\pm$.}}

As discussed in some detail, in the $x^\pm\in [0,2\pi]$ range horizons intersect the boundary in $(n_++1)(n_-+1)$ distinct regions, where $n_+, n_-$ are respectively number of zeros of $\psi$ and $\phi$ functions. Moreover, we showed that the inner and outer bifurcation curves $\Sigma_\pm$ also have the same number of disconnected pieces. The surface gravity $\kappa$, angular velocity $\Omega_{\cal H}$ and length $A_{\cal H}$ of the outer horizon is given in terms of the Floquet indices of the $\psi$ and $\phi$ solutions, ${\mathcal T}_\pm$, as
\be
\frac 2\kappa=\frac1{{\mathcal T}_+}+\frac1{{\mathcal T}_-},\qquad \Omega_{\cal H}=\frac{{\mathcal T}_+-{\mathcal T}_-}{{\mathcal T}_++{\mathcal T}_-},\qquad A_{\cal H}=2\pi\ell({\mathcal T}_++{\mathcal T}_-).
\ee

There is a closely related construction for multi-BTZ geometries due to Brill \cite{Brill}. This construction is based on the fact that constant time slice for AdS$_3$ has the line element $ds^2=(dz^2+dx^2)/z^2$ which is nothing but a 2d hyperboloid $H^2$. The latter may also be viewed as the Poincar\'e disk. It is well known that one may construct all 2d surfaces of genus $g\geq 2$ from orbifolds of $H^2$, by making identifications on it with a discrete subgroup of its $sl(2,\mathbb{R})$ isometry. It is then argued that if we add back the time direction, the 3d geometry we obtain in this way is a black hole geometry with multi-sector horizon and boundary. In a different terminology, one may cut through the maximally extended multi-BTZ geometry at the constant time slice passing through the horizon. In this way, one observes that the multi-BTZ geometry is indeed a wormhole with $g$ number of handles \cite{Wormhole,HEE-wormholes}.

\begin{figure}[p]
\vskip -2cm
\begin{center}
\includegraphics[scale=.50]{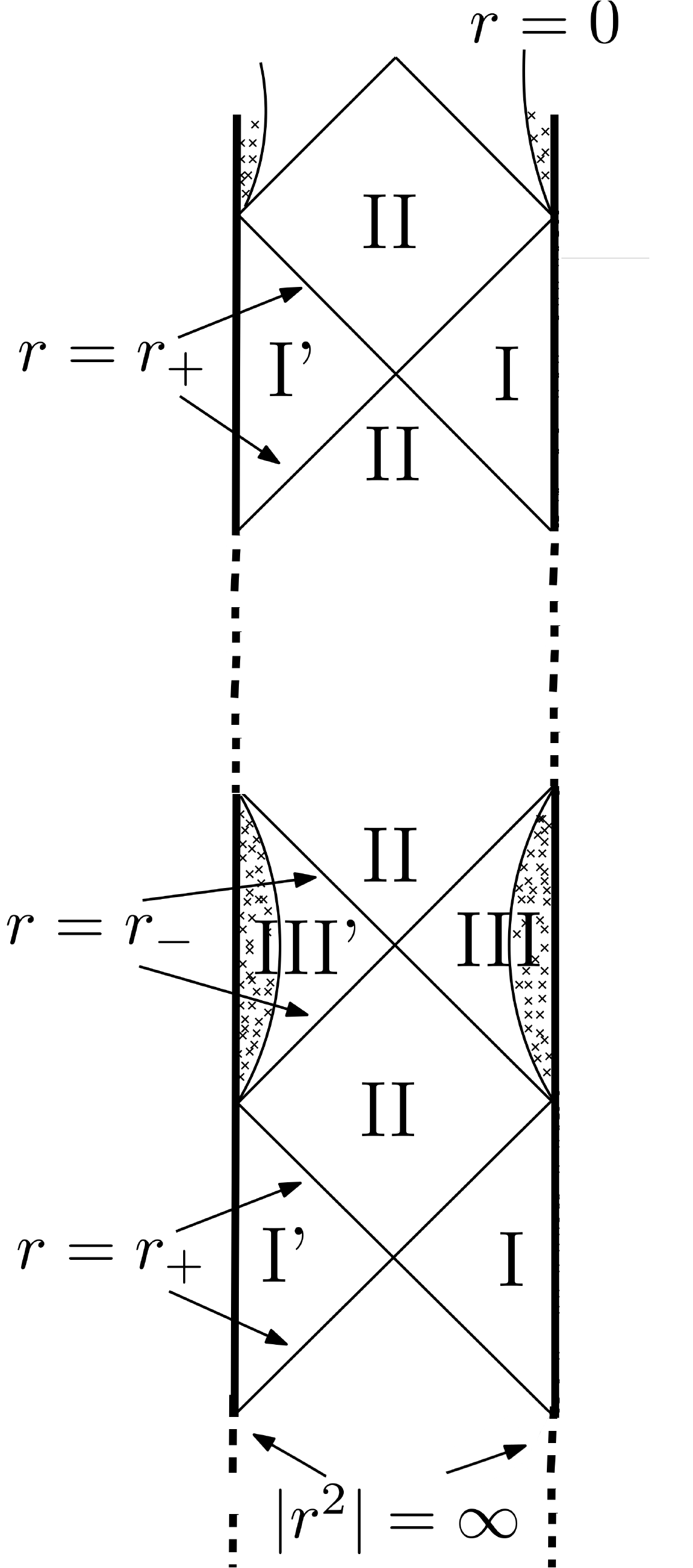}
\caption{
{Penrose diagram for generic Ba\~nados geometry. To draw this causal diagram we have used the analysis of the simpler case of BTZ  (\emph{cf.}) appendix A.1) and the analysis made in this section. 
In the Penrose diagram, as usual, we have suppressed a spacelike compact direction (here the one along $\Phi_+$ or $\Phi_-$ coordinates). Our discussions in this section reveal that  Penrose diagrams for generic $n_\pm$ are essentially the same as those of usual BTZ geometries discussed in \cite{BTZ}. However, we cut the regions which have CTCs  and we should make  appropriate identifications, and hence we remain with a geometry whose Penrose diagram is $(n_++1)(n_-+1)$ multiple repetition of that of a single BTZ geometry (Figure \ref{rho-vs-r-BTZ}).}}
\label{Penrose-diagram}
\end{center}\end{figure}

In the ``Brill's diagrams'' \cite{Brill}, the construction is based upon time-direction suppressed 3d geometry. In our tables, we discuss the $x^\pm$ while suppressing radial direction $r$ and in Penrose diagram (see Figure \ref{Penrose-diagram}) we are suppressing the ``compact $\Phi$'' direction. In a sense these three are complementary to each other. Making a thorough comparison and working out the details of connection between our multi-sector geometries and those of Brill is postponed to upcoming works.

\section{Ba\~nados geometries and the associated conserved charges}\label{sec-3}

As was discussed in \cite{AdS3-charges-everywhere}, to a given Ba\~nados geometry one may associate two kind of charges, the Virasoro charges and charges associated to Killing vectors $\zeta_\pm$. It was shown in \cite{AdS3-charges-everywhere} that the usual Lee-Wald \cite{Lee-Wald} or Barnich-Brandt \cite{Barnich-Brandt} symplectic structure vanishes on-shell for Ba\~nados geometries and hence within the covariant phase space method (see e.g. \cite{Hajian:2015xlp} for a review) one can define \emph{symplectic symmetries}. That is, the surface charges could be defined by integration on a space-like one-dimensional curve  everywhere in the geometry and not necessarily at the boundary. 

It was argued in \cite{Hajian:2015xlp} that sympelctic symmetries as defined above, may be in either of  the two groups: The generator of the symmetry could be an exact symmetry, in our setting a Killing vector, or it can be a nontrivial diffeomorphism which is not an exact symmetry. The former may hence be called ``Symplectic Exact Symmetry'' (SES) and the latter ``Symplectic Non-exact Symmetry'' (SNS). In our example the SNS are generated by diffeomorphisms along the $\chi$ vector field \eqref{BG-preserving-diff} and the SES are generated by the Killings $\zeta_\pm$ \eqref{KV} \cite{AdS3-charges-everywhere}. 

In this section we give the expression for charges associated with SNS and SES. The first part of this section is essentially a review of \cite{AdS3-charges-everywhere}. However, the second part is a completion of \cite{AdS3-charges-everywhere} along the analysis of \cite{Hajian:2015xlp}.

\subsection{Charges associated with Symplectic Non-exact Symmetries (SNS)}\label{sec-3-1}

In the covariant phase space method, we first define charge variations associated with field variations/perturbations generated by a vector field, e.g. $\chi$, and then integrate these charge variations over a given path in the phase space to define the charge itself. The integrability condition, which is checked to hold in our case \cite{AdS3-charges-everywhere}, then guarantees that the charges do not depend on the integration path in the phase space. 

For the Ba\~nados geometry with $L_\pm$ functions, if we denote the charges associated with $\chi[\epsilon_+=e^{inx^+}, \epsilon_-=0]$ and $\chi[\epsilon_+=0, \epsilon_-=e^{inx^-}]$ (\emph{cf.} \eqref{BG-preserving-diff}) respectively by $L_n, \bar L_n$, their explicit form is \cite{AdS3-charges-everywhere}
\be
L_n[g]=\frac{\ell}{8\pi G} \oint dx^+ L_+(x^+) e^{in x^+}\,,\qquad \bar L_n[g]=\frac{\ell}{8\pi G} \oint dx^- L_-(x^-) e^{in x^-}\,.
\ee
These charges form two Virasoro algebras at the Brown-Henneaux central charge:
\be
[L_n,L_m]=(n-m) L_{n+m}+\frac{c}{12}\delta_{m+n} n^3\,,\quad [L_n,\bar L_m]=0,\quad [\bar L_n,\bar L_m]=(n-m) \bar L_{n+m}+\frac{c}{12}\delta_{m+n} n^3\,,
\ee
with
\be\label{cenral-charge}
c=\frac{3\ell}{2G},
\ee
where $G$ is the 3d Newton constant. Note that unlike the usual Brown-Henneaux case \cite{Brown-Henneaux}, (1) our charges are defined everywhere and not just close to the boundary; (2) the charges are defined around an arbitrary Ba\~nados solution and not just AdS$_3$ (in Poincar\'e patch). This latter brings two interesting features: first, the charges are associated with each Ba\~nados geometry and second, the charges become field dependent, their expression explicitly depends on the background functions $L_\pm$. Therefore, to obtain the algebra of charges one needs to consider an ``adjusted bracket'' which includes the change made the background functions when computing charge variations; see \cite{AdS3-charges-everywhere} for detailed discussions and analysis.

\subsection{Charges associated with Symplectic Exact Symmetries (SES)}\label{sec-3-2}

In \cite{AdS3-charges-everywhere}  it has been argued that one can associate following conserved charges to U(1)$_\pm$ Killing vectors of the geometry (\ref{generic-solutions}). Using the analysis of \cite{Hajian:2015xlp}, we obtain\footnote{{To simplify the notation and avoid cluttering here we use $\delta X$ for denoting what was called parametric variations in \cite{Hajian:2015xlp}. In the latter paper, this was denoted by $\hat{\delta} X$.}} 
\be\label{delta-Jpm}
\delta J_{\pm}=\frac{\ell \;}{8\pi G} \int_0^{2\pi}{\cal K}^0_{\pm} {\delta L_{\pm}}\; dx^\pm \;,
\ee
where using Floquet form of the solutions \eqref{psi-phi-Floquet} and \eqref{Kpm}, 
\be\label{cal-K0}
{\cal K}_{+}^0=2{\mathcal T}_+\cdot\psi_1(x^+;{\mathcal T}_+) \psi_2(x^+;{\mathcal T}_+),\, \qquad {\cal K}_{-}^0=2{\mathcal T}_-\cdot\phi_1(x^-;{\mathcal T}_-) \phi_2(x^-;{\mathcal T}_-)\, 
\ee
and 
\be
\delta L_\pm= \frac{\partial L_\pm}{\partial {\mathcal T}_\pm} \delta {\mathcal T}_\pm.
\ee
The numeric coefficients $2{\mathcal T}_\pm$ in \eqref{cal-K0} is twice the Floquet index \eqref{psi-phi-Floquet} and has been added recalling the discussions around \eqref{Xpm-b'dry-coord} in section \ref{sec-2-3-2}.

Using (\ref{Schrodinger}) and \eqref{normalization}
we learn that 
\be\begin{split}
{\cal K}_{+}^0 {\delta L_{+}} =2{\mathcal {\mathcal T}}_{{+}}\left(\psi_2 \delta\psi_1''-\psi_2'' \delta\psi_1\right)
=2{\mathcal {\mathcal T}}_{{+}}\left(\psi_2 \delta\psi_1'-\psi_2'\delta\psi_1 \right)',
\end{split}\ee
and similarly for the  right-movers (the case labelled by minus index). In the above prime denotes derivative w.r.t. the argument $x^\pm$ and $\delta$ denotes variations w.r.t. the Floquet index ${\mathcal T}_\pm$. Therefore, 
\be
\delta J_{+}=\frac{\ell \;{\mathcal {\mathcal T}}_{+}}{4\pi G} \bigl[\psi_2 \delta\psi_1'-\psi_2'\delta\psi_1 \bigr]_0^{2\pi}\,,\qquad \delta J_{-}=\frac{\ell \;{\mathcal {\mathcal T}}_{{-}}}{4\pi G} \bigl[\phi_2 \delta\phi_1'-\phi_2'\delta\phi_1 \bigr]_0^{2\pi}.
\ee
One may now use the Floquet form \eqref{psi-phi-Floquet} and the normalization condition \eqref{normalization} to further simplify the above:
$$
\bigl[\psi_2 \delta\psi_1'-\psi_2'\delta\psi_1 \bigr]_0^{2\pi}=\bigl[x^+ \delta {\mathcal T}_+ \bigr]_0^{2\pi}=2\pi \delta {\mathcal T}_+,\qquad \bigl[\phi_2 \delta\phi_1'-\phi_2'\delta\phi_1 \bigr]_0^{2\pi}=\bigl[x^- \delta {\mathcal T}_- \bigr]_0^{2\pi}=2\pi \delta {\mathcal T}_-.
$$
To get the charges $J_\pm$ we need to integrate over $\delta J_\pm$ on path in the ${\mathcal T}_\pm$ space, i.e.
\be\label{charge-J-pm}
J_\pm=\frac{\ell}{2G}\int_{{\mathcal T}_{\pm 0}}^{{\mathcal T}_\pm} d\tilde {\mathcal T}_\pm {\tilde{\mathcal T}}_{{\pm}}=\frac{\ell}{4G} ({\mathcal T}_\pm^2-{\mathcal T}_{\pm 0}^2)
=\frac{c}{6} ({\mathcal T}_\pm^2-{\mathcal T}_{\pm 0}^2),\ee 
where ${\mathcal T}_{\pm 0}$ is a reference point which has zero $J_\pm$.  
As we will discuss in the next section two standard choices are ${\mathcal T}_{\pm 0}^2=-1/4$ (when $J_\pm$ are measured w.r.t. global AdS$_3$) or ${\mathcal T}_{\pm 0}^2=0$ when the reference point is chosen as AdS$_3$ in Poincar\'e patch (massless BTZ).

The very important point discussed in \cite{AdS3-charges-everywhere} is that the $J_\pm$ charges above commute with the Virasoro generators $L_n, \bar L_n$,
\be
[J_\pm, L_n]=[J_\pm,\bar L_n]=0,\qquad \forall n\in\mathbb{Z}.
\ee 

\paragraph{BTZ case.} For BTZ black hole solution, where $L_{\pm}={\mathcal T}_\pm^2$ and ${\mathcal T}_{\pm 0}^2=0$, we get
\be\label{BTZ-charges}
J_{\pm}=\frac{\ell}{4G}{\mathcal T}_{\pm}^2=\ell M_{BTZ}\pm J_{BTZ}\;.
\ee
Also, in this case the normalization factor $2{\mathcal T}_\pm$ is the  temperature in the left and right moving sectors.

\subsection{Entropy, the first law and Smarr relation for Ba\~nados geometries}\label{sec-3-3}

As in the standard Wald formulation \cite{Wald} the charge associated with the outer (event) Killing horizon is $S/2\pi$ where $S$ is the black hole entropy. In our case the charge associated with the Killing horizon generating vector fields 
$\zhpm$ is \cite{AdS3-charges-everywhere, Hajian:2015xlp}
\be\label{delta-S-def}
\frac{\delta S}{2\pi}=\frac{\ell}{8\pi G} \int (K_+\delta L_++ K_-\delta L_-).
\ee
Using the discussions of the previous subsection we learn that:
\be\label{first-law}
\frac{\delta S}{2\pi}=\beta_+\delta J_+ +\beta_-\delta J_-,\qquad \beta_\pm=\frac{1}{2{\mathcal T}_\pm}.
\ee
Note that $\delta$ in the above denotes variation in the Floquet indices ${\mathcal T}_\pm$.
Here $\beta_\pm$ is the inverse temperature associated with the left and right sectors.
The above is nothing but the first law for a generic Ba\~nados geometry. 
One can integrate \eqref{delta-S-def} in the parameter space, over ${\mathcal T}_\pm$ parameters, to obtain the Smarr relation for Ba\~nados geometries
\be\label{Banados-Smarr}
\frac{S}{2\pi}=\frac{\ell}{4G}({\mathcal T}_+ +{\mathcal T}_-)=2 (\beta_+J_++\beta_-J_-).
\ee
{We can write the entropy as a Cardy-type formula, using \eqref{charge-J-pm} \footnote{We thank D. Klemm for this remark.}
\be
S=2\pi\left(\sqrt{\frac{c(J_++J_{+0}) }{6}}+\sqrt{\frac{c(J_-+J_{-0})}{6}}\right),\qquad J_{\pm 0}=\frac{c}{6}{\mathcal T}_{0\pm}^2.
\ee} 
One may also compute the conserved charge  associated with the inner horizon generating Killing vector field. Straightforward computation, as performed above, leads to 
\be
\frac{S_{inner}}{2\pi}=\frac{\ell}{4G}({\mathcal T}_+ -{\mathcal T}_-).
\ee
This conserved charge may be written in terms of the other two conserved charges $J_\pm$. In particular one may note that
$$
S\cdot S_{inner}=\frac{\pi^2\ell}{G}(J_+-J_-).
$$

\section{Virasoro coadjoint orbits and their associated geometries}\label{sec-4}

It is well known that \emph{Virasoro algebra} 
\be\label{Virasoro-algebra}
[L_n,L_m]=(n-m) L_{n+m}+\frac{c}{12} n(n^2-1)\delta_{n+m},
\ee
is associated with the algebra of infinitesimal diffeomorphisms on a circle, $diff(S^1)$. That is, 
\be\label{infinitesimal}
x\to x+\xi(x),\qquad \xi(x+2\pi)=\xi(x),
\ee
where $[\xi(x)\partial_x, \rho(x)\partial_x]$ produces the Witt algebra, the Virasoro algebra without the central term.
The \emph{Virasoro group}, on the other hand, is associated with finite orientation-preserving coordinate transformations on the circle $Diff(S^1)$ which is generated through
\be\label{finite-coord-transf}
x\to h(x),\qquad h(x+2\pi)=h(x)+2\pi,\qquad h'>0.
\ee

Being infinite dimensional, representations of the Virasoro group, the corresponding ``Virasoro multiplets'', are also infinite dimensional. To be precise, we usually deal with ``coadjoint orbits'' instead of the representations of the  Virasoro group. The Virasoro coadjoint orbits are in one-to-one correspondence with Virasoro multiplets \cite{Kirillov, Segal, Witten-88}. 
The elements in the Virasoro coadjoint orbits are of course specified by periodic functions on the circle. However, given an arbitrary function $f(x)$ on $S^1$ there are elements in $Diff(S^1)$ under which $f$ does not change. One may then use this fact to distinguish functions (elements) which belong to the same or different orbits. Explicitly, let us recall that from the form of algebra \eqref{Virasoro-algebra} under the transformation \eqref{infinitesimal}, 
\be
\delta_\xi f=\xi'''-4f\xi'-2f'\xi,
\ee where the third order derivative appears due to the presence of the central term. \footnote{As is implied by the AdS/CFT correspondence the $\delta_\xi f$ is how the energy momentum tensor of a 2d CFT, $f$, transforms under conformal transformations.} Therefore, all the functions in the same orbit are generated by the solutions to $\delta_\xi f=0$, the stabilizer equation \cite{Witten-88,Kirillov}:
\be\label{stabilizer}
\xi'''-4f\xi'-2f'\xi=0.
\ee
The orbits associated with function $f(x)$, ${\cal O}_f$ are then \cite{Witten-88,  Kirillov, Segal}
\be\label{orbit-def}
{\cal O}_f= Diff(S^1)/T_\xi[f],
\ee
where $T_\xi[f]$ is the subgroup of $Diff(S^1)$ generated through the periodic solution(s) to stabilizer equation \eqref{stabilizer}; note that only periodic $\xi(x)$ can be ``exponentiated'' to give an element of the Virasoro group. 
So, the problem of classification of Virasoro coadjoint orbits reduces to classifying periodic solutions of the stabilizer equation \eqref{stabilizer}. This classification is well established and standard references on the topic are \cite{ Witten-88, Kirillov, Segal}, however, we will use the method based on $SL(2,\mathbb{R})$ monodromy  discussed in \cite{Balog} and follow its notations.

Before moving further, we make the first remarkable correspondence between the analysis of orbits and the Ba\~nados geometries: the stabilizer equation \eqref{stabilizer} is exactly the same equation which appeared in the analysis of Killing vectors of Ba\~nados metrics and the group $T_\xi[f]$ is nothing but the group of global isometries of Ba\~nados metrics with a given $f$. 

The rest of the analysis of solutions to \eqref{stabilizer} goes as discussed in the previous section, through the Hill's equation
\be\label{Hill-eq}
\psi''-f(x)\psi=0,\qquad f(x+2\pi)=f(x)\,,
\ee
where $x\in [0,2\pi]$ parametrize a circle of unit radius. It is straightforward to check that upon the coordinate transformation in Virasoro group \eqref{finite-coord-transf}, the pair $(\psi(x), f(x))$ in \eqref{Hill-eq} transform to $(\tilde{\psi}(x),\tilde f(x))$ where \cite{Balog}
\be\label{conf-transf}
\begin{split}
f(x)&\to \tilde f(x)=h'^2 f(h(x))- S(h;x) , \\
\psi &\to \tilde\psi(x)=\frac{1}{\sqrt{h'}} \psi(h(x)) \,,
\end{split}
\ee
where $S(h;x)$ is the Schwartz derivative
\be
S(h;x)=\frac{h'''}{2h'}-\frac{3h''^2}{4h'^2}.
\ee

Each coadjoint orbit will hence be specified by a ``representative character $f(x)$,'' the $\psi$ and the corresponding ``conformal descendants'' (constructed through \eqref{conf-transf}). The ``little group'' $T_\xi[f]$ by which we mod out the $Diff(S^1)$, is generically generated by $\psi_1\psi_2$, where $\psi_i$ are the two linearly independent solutions to Hill's equation \eqref{Hill-eq} in the Floquet form \eqref{psi-phi-Floquet}. However, in special case of $f=-n^2/4,\ n\in \mathbb{Z}$ Hill's equation has three periodic solutions (with $2\pi/n$ periodicity). In these cases $T_\xi[f]$ is $n$-fold cover of $PLS(2,R)$, $PLS^{(n)}(2,R)$ \cite{Balog}.\footnote{{As discussed the next subsection, the $f(x)=0$ is also special as $\psi_1, \psi_2$ do not strictly follow Floquet form.}}

As reviewed in \cite{Balog}, one can recognize two general class of such coadjoint orbits: those with a constant representative and the other with $x$-dependent representative.

\subsection{Constant representative coadjoint orbits}\label{sec-4-1}

There are four class of such orbits:
\begin{itemize}
\item \emph{Exceptional orbits} $E_n$, with representative:
\be\label{En-psi}
f_n=-\frac{n^2}{4}\,,\quad \psi_n=\sqrt{\frac{2}{n}} \sin\frac{nx}{2},\ \sqrt{\frac{2}{n}}\cos\frac{nx}{2},\qquad n\in\mathbb{Z}^+\,.
\ee
One may write the $\psi$'s in another Floquet form as
\be\label{En-psi-2}
\psi_n=\sqrt{\frac{1}{in}}e^{inx/2},\ \sqrt{\frac{1}{in}}e^{-inx/2}.
\ee
\item \emph{Elliptic orbits} $C(\nu)$, with
\be\label{Cnu-orbits}
f_\nu=-\frac{\nu^2}{4}\,,\quad \psi_\nu=\sqrt{\frac{2}{\nu}}\sin\frac{\nu x}{2},\ \sqrt{\frac{2}{\nu}}\cos\frac{\nu x}{2},\qquad \nu\notin\mathbb{Z}^+\,,\ee
or in the Floquet form
$$
\psi_\nu=\sqrt{\frac{1}{i\nu}}e^{i\nu x/2},\ \sqrt{\frac{1}{i\nu}}e^{-i\nu x/2}.
$$
Note that there is no overlap between exceptional and elliptic orbits.
\vskip 2mm
\item \emph{Zeroth Hyperbolic orbits} $B_0(b)$, with
\be\label{B0-b-orbits}
f_b=b^2\,,\quad \psi_b=\sqrt{\frac{1}{{2b}}}\; e^{bx},\ \sqrt{ \frac{1}{{2b}}}\;e^{-bx},\,\qquad b\in\mathbb{R}^+.
\ee
\item \emph{Zeroth order parabolic orbit} $P_0^+$,
\be\label{parabolic0+-orbits}
f=0\,,\qquad \psi=\frac{x}{\sqrt{2\pi}}\,,\ {\sqrt{2\pi}}.
\ee
\end{itemize}
Some comments about constant representative orbits are in order:
\bn
\item For all of the above orbits the Floquet index is either a real number (for hyperbolic orbits $B_0(b)$) or a pure imaginary number for exceptional or elliptic orbits $C_\nu$. 
\item The exceptional case $E_n$ is special as it has two Floquet forms, as written in \eqref{En-psi} and \eqref{En-psi-2}. The Floquet index in these two cases is either zero or $in$. 
\item The parabolic case is also special in the sense that the corresponding $\psi$'s are not in strict Floquet form, the $\psi_1$ is ``quasi-periodic'', $\psi_1(x+4\pi)=\psi_1(x)+2\psi_2(x)$. The generator of ``periodic'' Killing is $\psi_2^2$ (instead of $\psi_1\psi_2$). 
\item  The $b=0$ hyperbolic orbit overlaps with the $n=0$ exceptional orbit and both have $f=0$. However, this is still different from the parabolic orbit $P^+_0$ which again has $f=0$.
\item  Note that the above functions are for the ``representative'' of the orbit. A generic element in the orbit may be constructed from these upon the action \eqref{conf-transf} with \eqref{finite-coord-transf}. For generic element in the orbit, hence, the function $f$ is not a constant.
\en

\subsection{Non-constant representative coadjoint orbits}\label{sec-4-2}
There are two such orbits, parabolic ones $P_n^\pm$ and hyperbolic ones $B_n(b)$:
\begin{itemize}
\item \emph{Parabolic orbits} $P_n^\pm$, with
\be\begin{split}
f_n^\pm &=\frac{n^2}{2H_n}-\frac{3n^2(1\pm\frac{1}{2\pi})}{4H_n^2}\,,\\ 
\psi_n= \frac{1}{\sqrt{H_n}}&\left(\pm\frac{x}{2\pi}\sin\frac{nx}{2}-\frac{2}{n}\cos\frac{nx}{2}\right)\,,\ \frac{1}{\sqrt{H_n}}\sin\frac{nx}{2}\,,\qquad n\in\mathbb{N}\,.
\end{split}
\ee
where
\be
H_n(x)=1\pm\frac{1}{2\pi}\sin^2\frac{nx}{2}.
\ee
As one can explicitly see, $\psi$'s are not in standard Floquet form, and as the parabolic orbit $P_0^+$,  $\psi_2$ is periodic and $\psi_1$ is quasi-periodic,
$$
\psi_1(x+4\pi)=\psi_1(x)\pm 2\psi_2(x).
$$
The generator of ``periodic'' Killing is $\psi_2^2$ (instead of $\psi_1\psi_2$).
One may easily observe that $H_n(x)=H_{n=1}(nx),\ f_n^\pm(x)=n^2 f_{n=1}^\pm(nx)$. Moreover, $f_n^\pm(x)+n^2/4$ is not a positive definite function.
\vskip 2mm

\item \emph{Hyperbolic orbits} $B_n(b)$, with
\be
\begin{split}
f_{n,b}=b^2+\frac{b^2+4n^2}{2F}-&\frac{3n^2}{4F^2}\,,\quad \\
\psi_{n,b}=\frac{e^{bx}}{\sqrt{F(x)}}\sqrt{\frac{2}{n}}\left(\frac{b}{n} \cos\frac{n x}{2}+\sin\frac{nx}{2}\right),\ & \frac{e^{-bx}}{\sqrt{F(x)}}\sqrt{\frac{2}{n}} \cos\frac{n x}{2},\,\qquad b\in\mathbb{R}^+\,,n\in\mathbb{N},
\end{split}\ee
where
\be
F_{n,b}(x)=\cos^2\frac{nx}{2}+\left(\sin\frac{nx}{2}+\frac{2b}{n}\cos\frac{nx}{2}\right)^2\,.
\ee
One may check that 
\be\label{n,b/n}
F_{n,b}(x)=F_{n=1,b/n}(nx),\ f_{n,b}(x)=n^2 f_{n=1,b/n}(nx),\qquad \psi_{n,b}(x)=\sqrt{\frac{2}{n}} \psi_{n=1,b/n}(nx).
\ee 
Moreover, one may check that  $f_{n,b}+n^2/4$ is not a positive definite function. 
\end{itemize}

Some comments and points about the nonconstant representatives are in order:
\begin{enumerate}
\item
The $n=0$  hyperbolic and parabolic orbits cannot be obtained from the above nonconstant representative orbits  by setting $n=0$. 
\item The character function $f_n$ of the above orbits is a function with $2\pi/n$ periodicity.
\item The function $f_n$ for nonconstant representatives can become negative. Integral of $f_n+1/4$ which is often called ``energy'' \cite{Balog} has a negative value, except for the $P^-_{1}$ orbit. 
For hyperbolic orbits this ``energy'' is unbounded from below (as a functions of $b$). One may show that this ``energy'' does not have definite sign either, for the descendants of the representative.
\item As the equations above indicate, the representative element of  $n>1$ orbits may be expressed through $n=1$ ones though with replacing $x$ with $nx$. In particular, for the hyperbolic ones one should also replace $b$ parameter with $b/n$. 
\item A generic element in these orbits can be obtained from the above ``representatives'' upon the action \eqref{conf-transf} with \eqref{finite-coord-transf}. For generic element of the orbit then the character function $f$ or $\psi$'s are just $4\pi$ periodic (and not $4\pi/n$).
\item One may readily see that the hyperbolic orbits $B_n(b)$ and the exceptional orbits $E_n$ overlap at $b=0$.
\item $n$ is in fact determining the number of zeros of $\psi$ functions in $[0,2\pi]$ range.
\end{enumerate}

\subsection{Orbit invariant quantities}\label{sec-4-3}

As discussed, a Virasoro coadjoint orbit is specified by a representative function and a character, $\psi,\ f$, and then elements in the orbit, the descendants, are constructed from this upon the action \eqref{conf-transf}. So, each orbit consists of infinitely many (countable though, because $h(x)$ is a periodic function) functions/states. Since we build the orbit from the representative, one would expect that the parameters specifying the representative functions should be readable from any element in the orbit (and not just the representative of the orbit). Explicitly, there should be some ``orbit invariant'' charges and quantities. 

In section \ref{sec-3}, we have in fact laid the basic ground for specifying these charges and quantities:
any element in an orbit is specified by two kind of charges: the Virasoro charges (specified by combination of Virasoro generators L$_n$'s, $\Pi_{\{n_k\}} L_{-{n_k}}$) and the $J$ charges, which specify the representative. These two charges commute with each other. Our construction of the $J$ charges in section \ref{sec-3-2} makes it explicit that this charge is an orbit invariant quantity.  

Now that we have discussed the orbit classification, we can be more explicit about these charges. As we see in general any orbit is specified by a discrete integer label $n$ and/or a continuous label $b$ (for hyperbolic orbits) and $\nu$ (for elliptic orbits). By construction, the charge $J$ can only be associated with the continuous label on the orbit. The reason is that the method discussed in \cite{Hajian:2015xlp}, which is reviewed and used in section \ref{sec-3-2}, is suited for computing \emph{charge variations} within a given class of solutions with exact symmetry. This means that within a class of given orbits, e.g. the hyperbolic orbit with a given $n$, the orbits may be uniquely specified by the continuous label. 
This continuous parameter is related to the conserved charge $J$, as given in \eqref{charge-J-pm}. We note that the parameter ${\mathcal T}$ which specifies the charge $J$ is the Floquet index  defined as $exp(4\pi {\mathcal T})=\psi_1(x+4\pi)/\psi_1(x)$. With this definition, and recalling \eqref{conf-transf}, one immediately sees that ${\mathcal T}$ is orbit invariant, because
\be
e^{4\pi{\mathcal T}}=\frac{\tilde\psi(x+4\pi)}{\tilde\psi(x)}=\frac{\psi(h(x+4\pi))/\sqrt{h'}}{\psi(h(x))/\sqrt{h'}}=\frac{\psi(x+4\pi)}{\psi(x)}.
\ee

The discrete label on the orbits is also an orbit invariant quantity. This index is given by the number of zeros of $\psi_1$ or $\psi_2$ functions in the $[0,2\pi]$ range. To see the orbit invariance of this label, we recall \eqref{conf-transf} which states that the $\psi$ function of any two states in the same orbit are related to the representative element as
\be
\tilde\psi (x)=\frac{1}{\sqrt{h(x)}}\psi(h(x)),\qquad h(x+2\pi)=h(x)+2\pi,\quad h'>0.
\ee
The above clearly shows that number of zeros of $\tilde\psi$ and $\psi$ are the same.  We note that the $X^{\pm}$ coordinate define in (\ref{Xpm-b'dry-coord}) are also orbit invariant.  

Unlike the continuous label, the Floquet index, there is no ``Noether-type'' conserved charges associated with the discrete label $n$. There are however, topological charges (invariants) related to it.

\subsection{Ba\~nados geometries/Virasoro orbits correspondence}\label{sec-4-4}

With the discussions and analysis of earlier sections, we are now ready to match the Virasoro coadjoint orbits on the left and right sectors and the Ba\~nados geometries. A discussion on this was presented in section 4 of \cite{AdS3-charges-everywhere} our main addition to that list is the cases involving generic hyperbolic orbits.\footnote{Similar analysis and the geometric picture associated with the BMS group has also been discussed in some papers, e.g. \cite{BMS-orbits}.}  In general, depending on $L_+$ and $L_-$ functions the Ba\~nados geometry will be in $\mathbf{{\cal O}}_+\otimes \mathbf{{\cal O}}_-$ orbit, where $\mathbf{{\cal O}}_\pm$ denote the orbits associated with left and right sectors. Here we will mainly consider cases where the left and right sectors are both from the same class of orbits. Some of the ``mixed cases'' has been discussed in the appendix A.3. 

{Note that geometries in the same orbits, the descendants, are related to each other by a specific class of diffeomorphisms generated by vector field $\chi$ \eqref{BG-preserving-diff}. Being diffeomorphic to each other, the geometries in the same orbit  share the same causal, boundary and horizon structure, and are of course described with the same $J_\pm$ charges and the same horizon temperatures ${\mathcal T}_{\pm}$.}

\paragraph{Constant representative cases}
\begin{itemize}
\item $E_{n_+}\otimes E_{n_-}$ orbits. These are geometries which are descendant of $n_\pm$-fold covers of AdS$_3$ with 
$PLS^{(n_+)}(2,\mathbb{R})\times PLS^{(n_-)}(2,\mathbb{R})$ isometry.  We note that this is the symmetry (isometry) group of the representative of the orbit and the ``descendants'' generically have only $U(1)_+\times U(1)_-$ global isometry. Note that while the representatives have $L_\pm=-n^2_\pm/4$, geometries in this family  do not generically have a constant character function $L_\pm$. Geometries in this orbit are horizon-free and their global and boundary structure is like $N$-fold cover of AdS$_3$ (see appendix A.3 for more details).  The global AdS$_3$ is the representative element of the $n_\pm=1$ orbit.  
\vskip 2mm
\item $C(\nu_+)\otimes C(\nu_-)$ orbits include geometries which are descendants of particles of given mass and angular momentum on the $N$-fold covers of AdS$_3$ (see appendix A.3). These geometries do not have horizon and are not black holes.
\vskip 2mm
\item $B_0(b_+)\otimes B_0(b_-)$ orbits. These geometries include BTZ black holes and their (conformal) descendants. We stress that all the geometries in this class (with a given $b_\pm$) have the same $J_\pm$ charges. Unlike the usual lore, as we have discussed the correct charge assignment to these geometries is $J_\pm$ with the ``energy'' in the left and right sectors and not $L_0$ and $\bar L_0$.
As reviewed in the Appendix, the geometry corresponding to the representative of the orbit, has constant, positive $L_\pm$, is the usual BTZ geometry. The other geometries descending from this constant $L$ ones, are uniquely specified
by their Virasoro charges, ``Virasoro hairs'', while sharing the same causal, boundary and horizon, structure as their BTZ parent geometry. All these geometries have the same horizon area, surface gravity and horizon angular velocity. 
\vskip 2mm
\item $P^+_{0}\otimes P^+_{0}$ orbit. The representative of this orbit corresponds to null-selfdual AdS$_3$ orbifold 
\cite{Null-self-dual} and the other states in this orbit have $J_\pm=0$.  This geometries may be obtained in the near horizon limit of geometry in the orbit of massless BTZ black holes \cite{massless-BTZ}. The global isometry group in this class is $SL(2,\mathbb{R})\times SL(2,\mathbb{R})$, where the two $U(1)$ factors in $SL(2,\mathbb{R})$'s can be noncompact.
\vskip 2mm
\item $P^+_{0}\otimes B_0(b)$ orbits. The representative of these geometries is AdS$_3$ selfdual orbifold \cite{Self-dual-orbifold}. Geometries in this class of orbits are not black holes (do not have event horizon). Nonetheless, they may be obtained as the near horizon limit of extremal BTZ \cite{AdS3-charges-everywhere, DLCQ}. The geometries in this class have four global Killing vectors, forming an $SL(2,\mathbb{R})\times U(1)$ isometry group. Nonetheless, the $U(1)\in  SL(2,\mathbb{R})$ is noncompact \cite{AdS3-charges-everywhere}.\footnote{We comment that geometries in the $E_{1}\otimes B_0(b)$ orbits, too, have four global Killings forming an $SL(2,\mathbb{R})\times U(1)$ isometry group, where the two $U(1)$'s here are compact. These geometries are not black holes either.}  
\vskip 2mm
\item $P^+_{0}\otimes C(\nu)$ orbits with $0<\nu<1$. These orbits show ``chiral particles'' (those with equal mass and angular momentum) on AdS$_3$ and their descendants. 
\end{itemize}

\paragraph{Geometry of hyperbolic orbits, $B_{n_+}(b_+)\otimes B_{n_-}(b_-)$.}
As it stems from the discussions of section \ref{sec-3}, these are geometries which in general have inner and outer horizons and their boundary has some disconnected pieces. As a general, but rough, picture one may consider an $N$-fold cover of AdS$_3$ and perform the same construction of BTZ black holes \cite{BTZ} on each patch of this $N$-fold cover, such that all the geometries have the same horizon temperatures ${\mathcal T}_{\pm}$. The horizon and boundary contains an $N$-fold copy and when one starts from a given patch on the boundary, s/he can only access a specific patch of the horizon (and the geometry). This is the picture we have for multi-BTZ black holes in Brill's construction \cite{Brill}.
In the Euclidean version the boundary is a 2d surface with $(n_++1)(n_-+1)$-handles. Irrespective of which patch at the boundary we look into the bulk, one would see exactly the same geometry, with the same mass and angular momentum.

As we discussed, one can associate positive, definite charges $J_\pm$ to these geometries, while the integral of the character $f_{n,b}$ (which has been called ``energy'' in the literature, e.g. see \cite{Balog}) is not necessarily positive, definite. We also note that for these orbits generic $n$ and $n=1$ cases are related as \eqref{n,b/n}. This, together with the relation for the entropy \eqref{Banados-Smarr}, may suggest that different patches of the horizon, while at the same temperature, carry an equal portion of the entropy, and that each patch has its own ``first law'' \eqref{first-law}.

\section{Concluding remarks and outlook}\label{sec-5}

In this work we elaborated further on the Ba\~nados geometries. This work was a continuation of \cite{Banados-Killings} and \cite{AdS3-charges-everywhere}. Below we provide a quick summary and a general picture arising from our analysis.
\bn
\item Different probes can access different kind of information from the geometry. The ``classical'' probes, like geodesics, have only access to ``classical, geometric'' information. These geometric information are ``orbit invariant'' and include geodesic distances, charges associated with the exact Killing symmetries $J_\pm$, causal and boundary structure. Classical observers are blind to Virasoro charges, ``Virasoro hair''. 
\item
The Virasoro charges are semi-classical ones, they are given by ``surface integrals,'' integrals over codimension two spacelike curves (that is, integrals over one-dimensional curves in our 3d case). This information is not available to local classical probes. 
\item There is a one-to-one relation between two copies of Virasoro coadjoint orbits and Ba\~nados geometries, which are specified by two general holomorphic functions. All geometries in the same orbit share the same ``geometric'' information, while they can be distinguished by their ``Virasoro hairs''. The geometric information are \emph{orbit invariant}.
\item It is possible that at the level of the geometry we have extra requirements like absence of CTCs, which needs to be considered.
\item  Both the charges associated with exact symmetries $J_\pm$ and the Virasoro hairs are ``symplectic charges'' \cite{AdS3-charges-everywhere, Hajian:2015xlp}. That is, these charges may be defined by integration over any spacelike compact curve in the 3d spacetime; this curve need not be at the boundary of the space or at its horizon.
\item Our analysis suggests the following general picture: different geometries which are diffeomorphic to each other  share the same ``geometric information''. This is what we learn in the standard GR courses. However, there could be a measure-zero set of diffeomorphisms producing semi-classically different geometries, they may be distinguished by their other surface charges, ``semi-classical hairs''. The states sharing the same classical geometric information fall into orbits of the semi-classical symmetry algebra. This symmetry algebra is a symplectic symmetry of the phase space constituted from diffeomorphic but distinguishable, geometries. If the geometry we are dealing with is a black hole, then the geometries which share the same geometric information may be viewed as ``hairs'' on this black hole. This, we hope, provides a handle on the black hole microstate problem.  We have established this picture for AdS$_3$ case. Similar ideas have been worked through for near horizon extremal geometries \cite{CHSS}. We think this picture should be more general and applicable to any black hole. This picture which was summarized and sharpened in \cite{Residual} fits well with the recent ideas and analysis \cite{Soft-hair}. We hope to provide further evidence for this picture in more general settings.
\item Although we worked in a specific gauge, the Ba\~nados coordinate system, we believe the above picture is gauge independent. First explicit steps in this direction was taken in \cite{AdS3-charges-everywhere}, where it was shown that similar results hold in the Gaussian null cordinates (also known as the BMS gauge). 
\item At a perhaps more technical level, Ba\~nados geometries form a phase space. Elements in this phase space are classified by the Virasoro coadjoint orbits. One may hence use this picture to perform quantization of AdS$_3$ gravity.
We hope this viewpoint can shed further light on the question of AdS$_3$ gravity quantization, e.g. see \cite{Carlip, AdS3-Q-gravity} and references therein.
\en

\subsection*{Acknowledgement}

We would like to thank Glenn Barnich, Steve Carlip, Geoffrey Comp\`ere, Kamal Hajian, Dietmar Klemm,  Hai Lin,   Joan Sim\'{o}n, Wei Song and Jun-Bao Wu for discussions and useful comments on the draft. We also thank Dongsu Bak and Soo-Jong Rey for early collaboration. We especially thank Ali Seraj for the discussions about the context of appendix A.1. 
The work of M.M. Sh-J is supported in part by Allameh Tabatabaii Prize Grant of
Boniad Melli Nokhbegan of Iran, the SarAmadan grant of Iranian vice presidency in science and technology and the ICTP network project NET-68. M.M.Sh-J. is also acknowledging the ICTP Simons fellowship. We thank ICTP for the hospitality while this project completed.

\appendix
\section{More on constant representative orbits and their geometries }\label{Appen-Constant-L}

In this appendix we give more detailed discussions and computations on the constant $L_\pm$ geometries and the associated orbits. This class includes BTZ black holes and conic spaces (particles on AdS$_3$).

\subsection{More on geometry of constant $L_\pm$ cases}\label{appendix-A-1}
We discussed in section \ref{sec-4}, there are four class of solutions to the Hill's equation with a constant representative which are elliptic orbits, exceptional orbits and zeroth of parabolic and hyperbolic orbits. This gives totally 10 classes of geometry solutions. Four of them  correspond to cases with the left and right sectors in the same orbit and others to the mixed orbits. The number of global Killing vector can be two, four or six. Two general Killing vectors are $\partial_+$ and $\partial_-$. Assuming two linearly independent combinations of these two Killing vectors denotes by $\zeta_{\pm}=\partial_+ \pm k \partial_-$ (with a real non-zero $k$). The norm of these vectors is given by
\be
|\zeta_{\pm}|^2=\mp \frac{(r^2\mp\ell^2kL_-)(kr^2\mp\ell^2L_+)}{r^2},\quad \zeta_+ \cdot  \zeta_-=\ell^2 (L_+-k^2L_-)\;.
\ee

Depending on the relative signs of $L_+$ and $L_-$, one can distinguish different possibilities.  When $L_+$ and $L_-$ have opposite signs,  the inner product does not vanish anywhere, while $|\zeta_{\pm}|$ vanish at  $r^2=\pm \frac{\ell^2L_+}{k},   r^2=\pm k\ell^2L_-$.  When they have similar relative sign, we may take $k^2=\frac{L_+}{L_-}$ leading to
\be\label{zeta-norm-const-L}
|\zeta_{\pm}|^2=\mp \sqrt{\frac{L_+}{L_-}}\frac{(r^2\mp\ell^2 \sqrt{L_+L_-})^2}{r^2},\quad \zeta_+ \cdot  \zeta_-=0,
\ee 
therefore $r^2=\pm\ell^2 \sqrt{L_+L_-}$ is \emph{{potentially}} the position of the bifurcate horizon.

When $L_{\pm}$ are constant, we can use the coordinate transformation
\be
\label{coortran}
\rho^2=\frac{\left(r^2+\ell^2  L_+\right) \left(r^2+\ell^2 L_-\right) }{r^2},\quad t=\frac{\ell}{2}(x^++x^-), \quad \varphi=\frac{1}{2}(x^+-x^-)\,,
\ee
and rewrite the metric in the ``BTZ form''
\be\label{generaladm}
ds^2=-F(\rho)dt^2 +\frac{d\rho^2}{F(\rho)} +\rho^2\left(d\varphi-N^{\varphi}dt\right)^2,
\ee
where
\be\label{fn}
F(\rho)=\frac{(\rho^2-\rho_+^2)(\rho^2-\rho_-^2)}{\ell^2 \rho^2}, \qquad N^{\varphi} = \frac{\rho_+\rho_-}{\ell \rho^2},
\ee
and
\be\label{BTZ-horizon-radii}
\rho_\pm=\ell(\sqrt{L_+}\pm \sqrt{L_-}).
\ee

In this coordinate system, \eqref{zeta-norm-const-L} takes the form
\be\label{zeta-norm-BTZ-coord}
|\zeta_{\pm}|^2=\mp \left(\frac{\rho_++\rho_-}{\rho_+-\rho_-}\right)\left(\rho^2-\rho^2_\pm\right)^2.
\ee
As we see for generic $L_\pm$ with arbitrary signs, $\rho_\pm$ need not be real-valued, and we hence need to consider the three cases of $L_\pm>0$, $L_\pm<0$ and $L_+ L_-<0$ separately.

\begin{figure}
\begin{center}
\includegraphics[scale=.5]{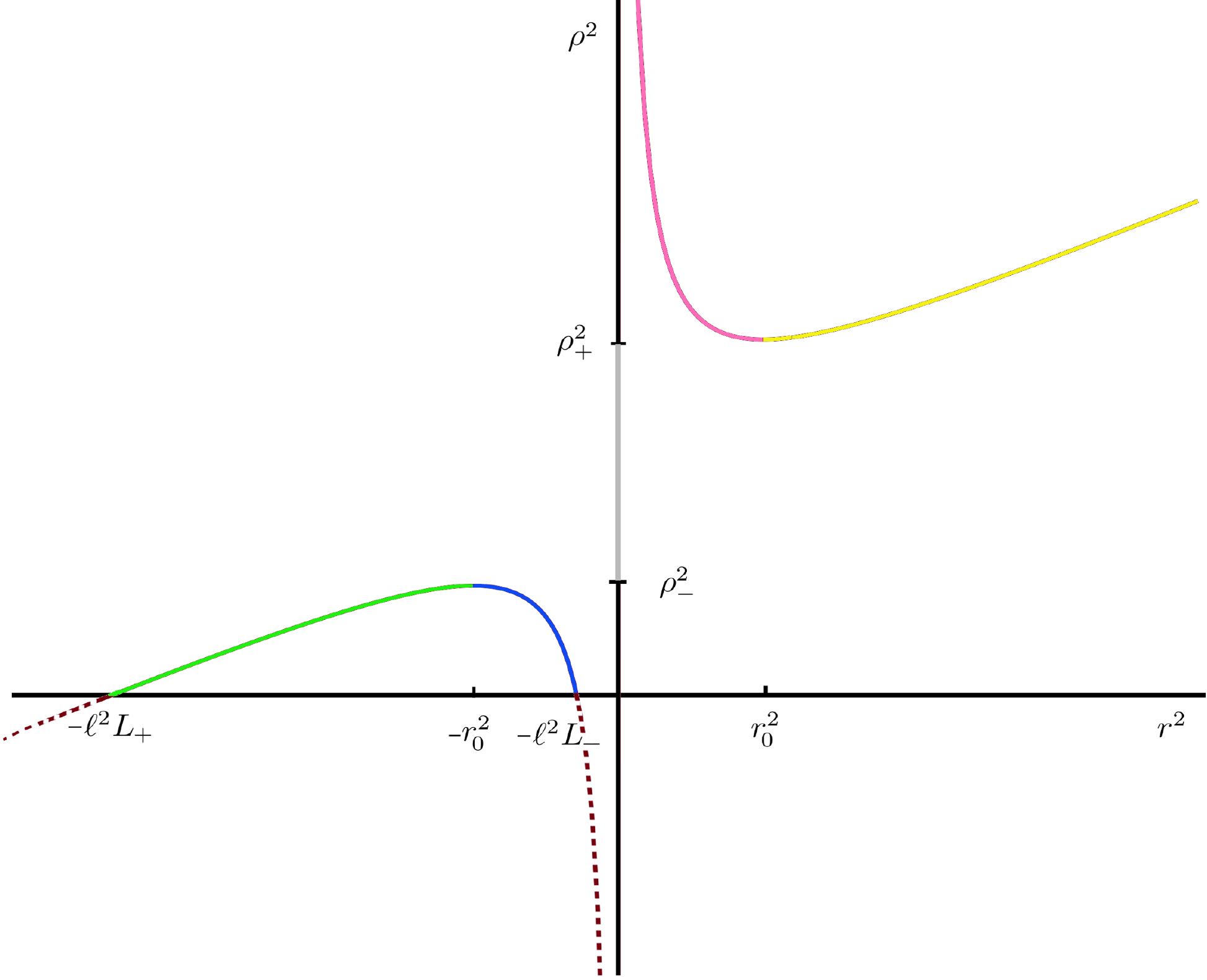}\qquad\includegraphics[scale=.45]{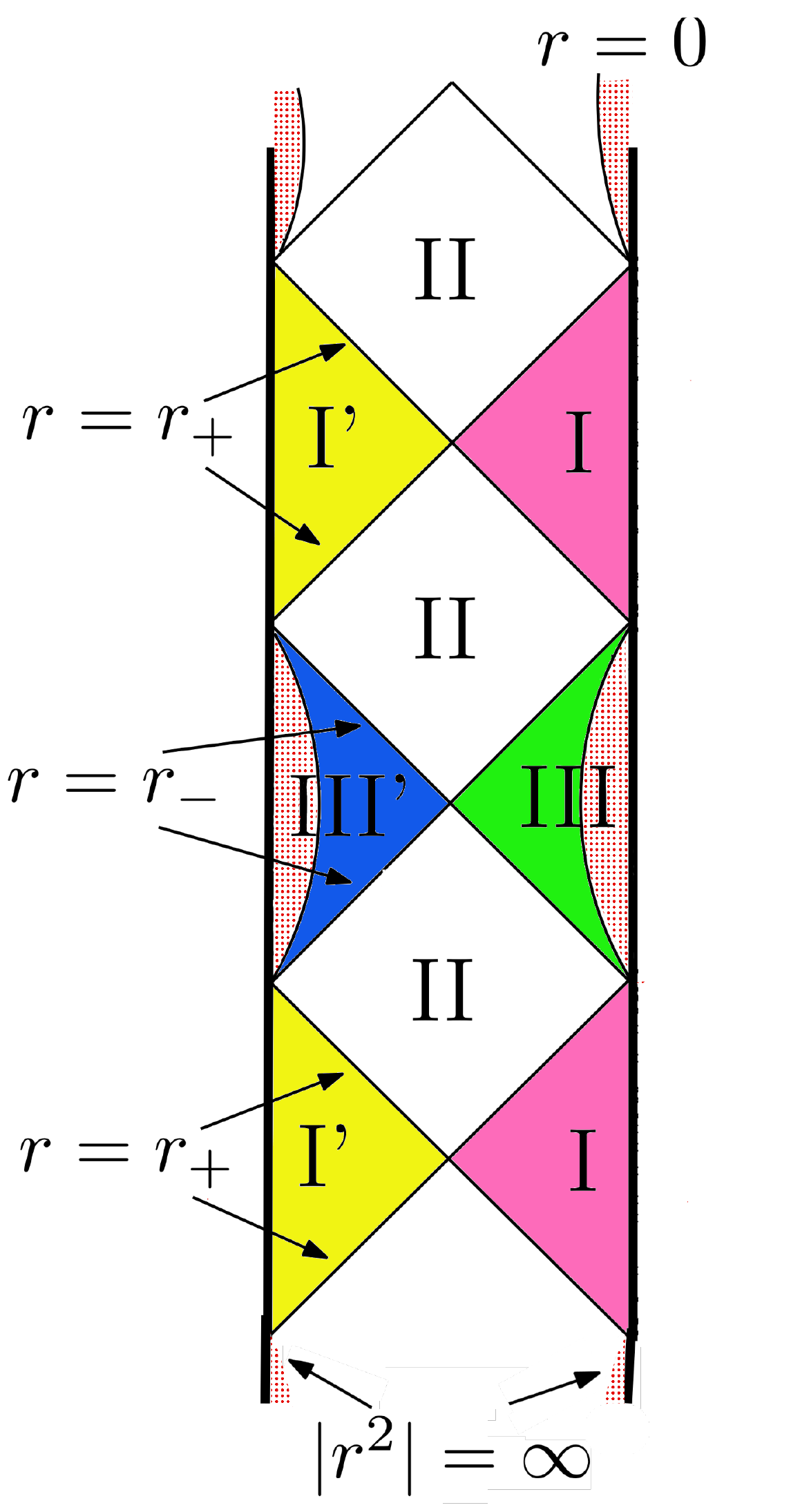}
\caption{Left: Coordinate transformation \eqref{coortran} plotted for the BTZ case of $L_+> L_->0$. Vertical axis denotes $\rho^2$ and horizontal axis  $r^2$. The (red) dotted line, where $\rho^2<0$, is the location of CTC.  Therefore, the CTC-free region in the Ba\~nados coordinate system is $-\ell^2 L_+ <r^2< -\ell^2 L_-$ and $r^2>0$. The region on vertical axis  in gray color, $\rho_-^2<\rho^2<\rho_+^2$, is not covered in the Ba\~nados coordinate system, while the other $\rho^2>0$ regions are covered twice.
The four regions $-\ell^2 L_+ <r^2\leq -r_0^2,\ -r^2_0\leq r^2<-\ell^2 L_-,\ 0<r^2\leq r_0^2$ and $r^2\geq r_0^2$ (with $r_0^2=\ell^2\sqrt{L_+L_-}$) which are also color-coded in the Left figure, correspond to the four regions, four diamonds, on the Penrose diagram (Right).\newline 
Right: Penrose diagram for the BTZ case of $L_+> L_->0$ \cite{BTZ, Loran}. The region II (which lies between the inner and outer horizons) is not covered in the Ba\~nados coordinate system. The regions I and I' respectively corresponds to $0<r^2\leq r_0^2$ and $r^2\geq r_0^2$ regions and the regions III and III' to $-\ell^2 L_+ <r^2\leq -r_0^2,\ -r^2_0\leq r^2<-\ell^2 L_-$.
The shaded regions are where we have CTC's and correspond to the (red) dotted regions in the Left figure. We have used the same color-coding in the Left and Right figures to indicate the range of $r^2$ coordinate. This figure shows how the Ba\~nados and BTZ coordinate systems are  complementary to each other.}
\label{rho-vs-r-BTZ}
\end{center}\end{figure}

\paragraph{$L_+, L_->0$, the BTZ case.}
In this case both $\rho_\pm$ are real-valued and are denoting the horizon radii. As \eqref{BTZ-horizon-radii} shows, $\rho_+$ ($\rho_-$) is the radius of the outer (inner) horizons. The event horizon is generated by $\zeta_{{\cal H}}$, where
$$
\zeta_{{\cal H}}=\partial_t+\Omega \partial_{\varphi}, \qquad \Omega=\frac{\rho_-}{\ell \rho_+}.
$$
The surface gravity is given by 
$$
\kappa=\frac{\rho_+^2-\rho_-^2}{\ell \rho_+}.
$$
To avoid the CTC we need to limit $\rho^2$ to $\rho^2>0$ region. Using \eqref{coortran} we can readily find where in the Ba\~nados radial coordinate $r^2$ is CTC-free. This has been depicted in Figure \ref{rho-vs-r-BTZ}.

The conserved charges $J_\pm$ are related to mass and angular momentum of BTZ black hole as in \eqref{BTZ-charges}. In this case the appropriate choice for ${\mathcal T}_{\pm 0}$ is zero, so that the massless BTZ has vanishing mass and angular momentum. 

\paragraph{$L_+, L_-<0$, the conic spaces.} In this case $\rho_\pm^2$ are negative and hence there are no Killing horizons ($|\zeta_\pm|=0$) in the region which there are no CTC. These geometries are hence horizon-free and are not black holes. These correspond to particles on AdS$_3$ (the spaces with conical defects) \cite{3d-solns}\footnote{See \cite{Raeymaekers:2014kea} for some discussion on quantization }. These particles are specified by mass $M$ and angular momentum $J$,
\be
\ell M\pm J=J_\pm=\frac{c}{6}(\frac14+L_\pm).
\ee
The above is nothing but \eqref{charge-J-pm} with ${\mathcal T}_{\pm 0}^2=-1/4$.

\begin{figure}[ht]
\begin{center}
\includegraphics[scale=.5]{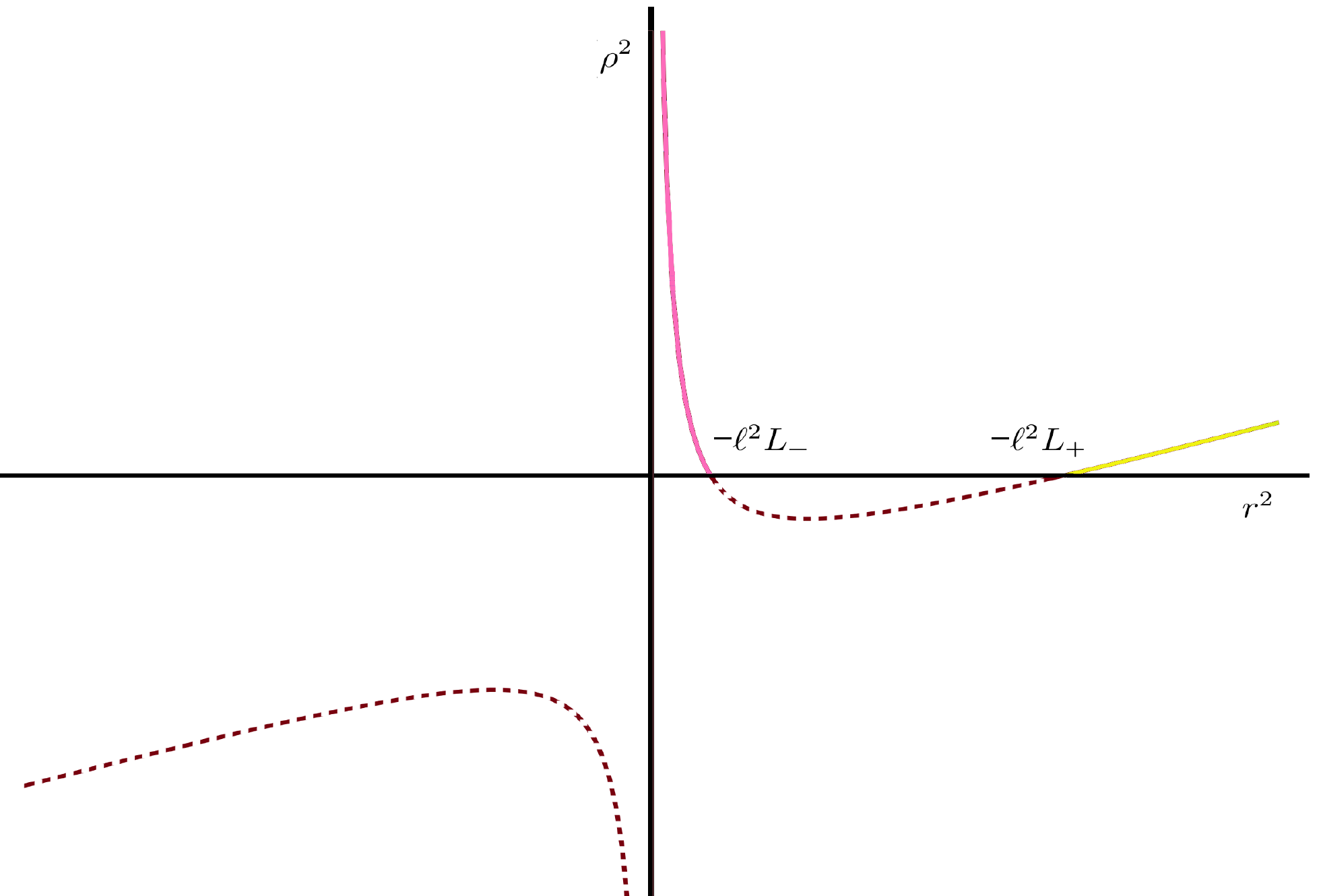}
\caption{ Coordinate transformation \eqref{coortran} plotted for the conic space with $L_+<L_- <0$. CTC region, where $\rho^2<0$, is denoted by (red) dotted curve. Extending coordinate to negative $r^2$ gives CTC. Positive values of $r^2$ with $r^2>-\ell^2L_+$ gives one cover denoted by yellow colour in above figure and $r^2<-\ell^2 L_-$ gives another cover which is denoted by by pink colour. Region $-\ell^2 L_-<r^2<-\ell^2L_+$ gives CTC. Therefore, in the CTC-free range there is no horizon. This is compatible with the fact that conic spaces correspond to particles on AdS$_3$ and not black holes.}
\label{rho-vs-r-conic}
\end{center}
\end{figure}

To write out metric explicitly, we may introduce $\tilde{\rho}_{\pm}^2=-{\rho}_{\pm}^2>0$. Then metric has the same form as (\ref{generaladm}) with
\be
F(\rho)=\frac{(\rho^2+\tilde{\rho}_+^2)(\rho^2+\tilde{\rho}_-^2)}{\ell^2\rho^2},\quad N^{\varphi}= -\frac{\tilde{\rho}_+\tilde{\rho}_-}{\ell \rho^2}\;.
\ee
To avoid the existence of CTC (see Figure \ref{rho-vs-r-conic}), we must restrict ourselves to $\rho^2>0$ region which is corresponding to
\be
r^2>0,\quad r^2 \notin (\ell^2 |L_-|, \ell^2 |L_+|)\;.
\ee

\paragraph{$L_+L_-<0$, the mixed case.} Without loss of generality we may choose $L_+>0, L_-<0$.

\begin{figure}[h]
\begin{center}
\includegraphics[scale=.5]{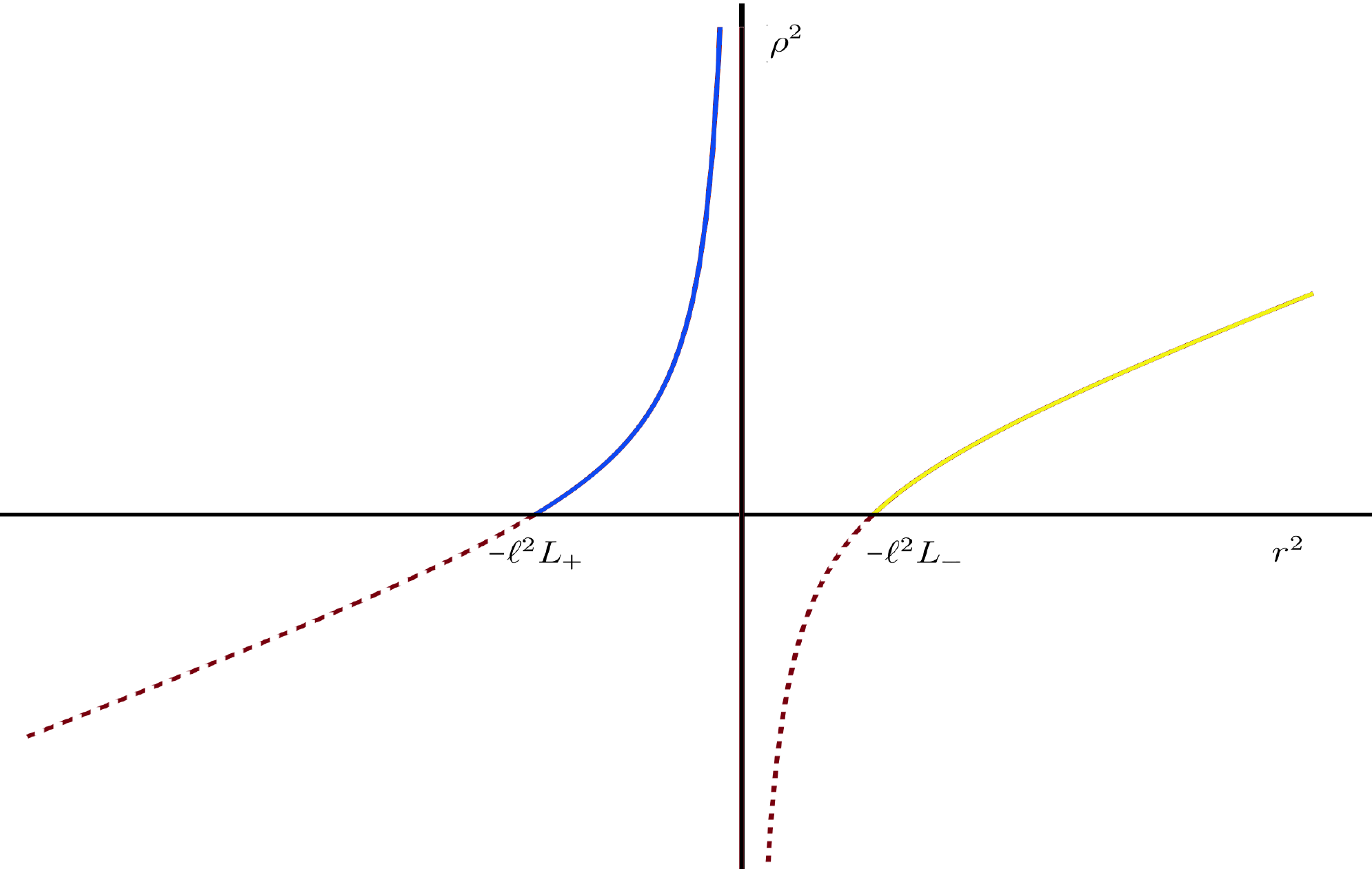}
\caption{ Coordinate transformation \eqref{coortran} plotted for $L_+>0$ and  $L_-<0$. The regions  $r^2>-\ell^2 L_-$ and $-\ell^2 L_+ <r^2<0$ are the CTC-free regions. The geometry does not have horizon in this region. The two CTC-free pieces both correspond to the same coordinate range $\rho^2>0$ in the BTZ-coordinate system.
}
\label{rho-vs-r-mixed}
\end{center}
\end{figure}In this case  $\rho_{\pm}$ turn out to be complex conjugate of each others and metric is given by \eqref{generaladm} with
\be\label{mixed-B-C-metric}
F(\rho)=\frac{\left[\rho^2-\ell^2(L_++L_-)\right]^2-4L_+L_-}{\ell^2\rho^2},\quad N^{\varphi}=\frac{\ell(L_++L_-)}{ \rho^2}\;.
\ee

\subsection{Geodesic motion on massless BTZ} Starting with AdS$_3$ solution which is corresponding to $L_+=L_-=0$. The metric can be written as
\be\label{ads3metric}
ds^2=\frac{\ell^2 d\rho^2}{\rho^2}-\rho^2 dy^+dy^-\;.
\ee
Geodesic equations are given by
\be\label{AdSgeodesy}
\ddot{\rho}-\frac{\dot{\rho}^2}{\rho}+\frac{r^3\dot{y}^+\dot{y}^-}{\ell^2}=0,\quad \ddot{y}^++\frac{2\dot{\rho}\dot{y}^+}{\rho}=0,\quad \ddot{y}^-+\frac{2\dot{\rho}\dot{y}^-}{\rho}=0\,.
\ee
where $dots$ denote derivative with respect to the proper time $\tau$. The norm of velocity is given by
\be \label{geodesynorm}
|\dot{\rho}\partial_{\rho} +\dot{y}^+\partial_++\dot{y}^-\partial_-|^2=\frac{\ell^2}{\rho^2}\dot{\rho}^2-\rho^2\dot{y}^+\dot{y}^-,
\ee
which is equal $1$ time-like geodesies and $0$ for null geodesies. This equation can be solved to find $\dot{\rho}$ in terms of $\dot{y}^{\pm}$ for the null and time like geodesies separately, then using equations (\ref{AdSgeodesy}) to solve $\dot{y}^{\pm}$.
\paragraph{Null geodesies.} Combining (\ref{geodesynorm}) and (\ref{AdSgeodesy}) for the null geodesy give us $\ddot{\rho}=0$, which gives $\rho=\frac{2}{\ell}\sqrt{p_+p_-} \tau$ (after a shift in $\tau$).   Using this solution we can solve $y^{\pm}$

\be
y^+-y_0^{+}=\frac{\ell^2}{2\tau p_+} ,\quad y^--y_0^{-}=\frac{\ell^2}{2\tau p_-} ,\qquad p_+p_-, \tau\geq 0\;.
\ee  

Zero angular velocity geodesics are corresponding to $p_+=p_-$. In this case
\be
y^+-y_0^+=y^--y_0^-=\frac{\ell}{\rho}\;.
\ee
\paragraph{Time-like geodesies.}
 For the time-like geodesic, by combining (\ref{geodesynorm}) and (\ref{AdSgeodesy}) we get $\ddot{\rho}-\ell^{-2} \rho=0$, which gives $\rho=p_+e^{\frac{\tau}{\ell}}-p_-e^{-\frac{\tau}{\ell}} $.   Using this solution we can solve $y^{\pm}$
 \be
 y^+-y^{0+}=\frac{\ell p_m e^{-\frac{\tau}{\ell}}}{p_+(p_+e^{\frac{\tau}{\ell}}-p_-e^{-\frac{\tau}{\ell}}) },\quad y^--y^{0-}=\frac{\ell  e^{-\frac{\tau}{\ell} }}{p_+e^{\frac{\tau}{\ell}}-p_-e^{-\frac{\tau}{\ell}}}\;.
 \ee
\subsection*{Null and time-like geodesies of BTZ black holes}
Next let us consider a BTZ black hole, i.e constant non-zero $L_{\pm}$s. Noting that in this case we have only two global Killing vectors $\partial_{\pm}$. Using the fact that $K.v=const.$ where $K$ is a Killing vector and $v$ is the velocity vector we get following equations
\be
\ell^2L_+\dot{x}^+-\frac{r^4+\ell^2L_+L_-}{2r^2}\dot{x}^-=p_+,\quad \ell^2L_-\dot{x}^--\frac{r^4+\ell^2L_+L_-}{2r^2}\dot{x}^+=p_-\;.
\ee
These equations give two components of the velocity vector.
\be
\dot{x}^{\pm}=-\frac{2r^2(r^4p_{\mp} + 2\ell^2 L_{\mp}p_{\pm} r^2 +\ell^4L_+L_-p_{\mp})}{(r^4-\ell^2L_+L_-)^2}.
\ee

The norm of velocity vector field is given by

\be
|v|^2=\frac{\ell^2 \dot{r}^2}{r^2}-\frac{4r^2(p_+r^2+\ell^2 L_+p_-)(p_-r^2+\ell^2 L_-p_+)}{(r^4-l^4L_+L_-)^2}=0,1\;. \quad (\mathrm{null,timelike})
\ee
Therefore we can read $\dot{r}$ for the null and time like geodesies. 

\subsection{More on geometries of constant representative orbits}\label{appendix-A-2}

As discussed in section \ref{sec-4}, Ba\~nados geometries are in one-to-one relation with Virasoro coadjoint orbits on the left and right sectors. Here we discuss geometries corresponding to orbits of constant representative in more details. 

\paragraph{Similar constant orbits.} As we discussed geometric properties, such as horizon and causal structure of all the solutions in the same orbit are the same. We will hence  only focus on geometry of the representative  element of the orbit.
\begin{itemize}

\item {\bf $E_{n_+}\otimes E_{n_-}$ orbits:}
In this case the representative element has  $L_{\pm}=-\frac{n_{\pm}^2}{4}$ with integer $n_{\pm}$. $n_+=n_-=1$ corresponds to the global AdS$_3$ geometry. For general $n_{\pm}$ once again we get 6 global Killing vectors. However the periodicities of Killing vectors are $\frac{2\pi}{n_{\pm}}$. If we scale coordinates with the least common multiple of $n_+$ and $n_-$ as follows
\be
r\rightarrow N r,\quad x^{\pm}\rightarrow N^{-1} x^{\pm},\qquad N=lcm(n_+,n_-),
\ee
the metric turns out to be
\be
ds^2=\frac{\ell^2dr^2}{r^2}-\left(rdx^+-\frac{\nu_-^2\ell^2}{4r}dx^-\right)\left(rdx^--\frac{\nu_+^2\ell^2}{4r}dx^+ \right),
\ee
where $\nu_{\pm}=\frac{n_{\pm}}{N}$.  

When $n_+=n_-=N$, $\nu_{\pm}=1$ and we get $n$-fold cover of AdS$_3$. Explicitly if we perform coordinate transformation (\ref{coortran}), we arrive at the following metric
\be
ds^2=-\frac{\rho^2+\ell^2}{\ell^2}dt^2+\ell^2 \frac{d\rho^2}{\rho^2+\ell^2}+\rho^2d\varphi^2,\qquad  \varphi\in [0,2\pi N]\;.
\ee
For general $n_+\neq n_-$ it is clear that $\nu_{\pm}<1$, and the metric is like \eqref{generaladm} with \eqref{fn}, 
\be
\rho_{\pm}^2=-\frac{\ell^2}{4} (\nu_+\pm \nu_-)^2,\quad \varphi\in[0,2\pi N]\;.
\ee
This metric is representing particles on $N$-fold cover of AdS$_3$, see below.

\item {\bf $C(\nu_+) \otimes C(\nu_-)$ orbits:} When $L_{\pm}=-\frac{\nu_{\pm}^2}{4}$ are negative constants with non-integer $\nu$, the corresponding geometry is a particle in AdS$_3$, if $\nu_{\pm}<1$. 
As discussed in appendix \ref{appendix-A-1} geometry does not have event horizon.

When one or both of $\nu_{\pm}$ are larger than one, we can rewrite the metric as described for $E_{n_+}\otimes E_{n_-}$  case above, with 
\be
\rho_{\pm}^2=-\frac{\ell^2}{4} (\tilde\nu_+\pm \tilde\nu_-)^2,\qquad \tilde\nu_\pm=\frac{\nu_\pm}{N},\qquad N=lcm([\nu_-],[\nu_+])\;.
\ee
As we see $\tilde{\nu}_{\pm}<1$.

\item {\bf $B_0(b_+) \otimes B_0(b_-)$ orbits:}
When both of $L_{\pm}$ are constant positive ($L_{\pm}=b_{\pm}^2>0$) the corresponding geometry is a BTZ black hole. This solutions is discussed in appendix \ref{appendix-A-1} in details.

\item {\bf $P_0^{+} \otimes P_0^{+}$ orbits:} 
The zeroth order of the parabolic orbit has $L=0$. This can also happen in the constant hyperbolic orbits with zero character. The corresponding geometry with $L_{\pm}=0$ in the parabolic sector is the null self-dual orbifold \cite{Null-self-dual}. The solution is identical to the near horizon limit of massless  BTZ black hole solution  \cite{massless-BTZ}. One should note that solutions to the Hill's equation in this case do not have  the  Floquet form and to construct the global Killing vectors we use only the constant solution.   We comment that for the cases with parabolic orbit $P_0^+$ we have the possibility of having one or three global Killing vectors for each left or right sector. When we have three Killing vectors $U(1) \in SL(2,\mathbb{R})$ isometry is not compact. 

\end{itemize}

\paragraph{Mixed constant orbits.}
There are six possible combination of the left and right sectors with different constant orbits. 
\begin{itemize}
\item {\bf $ E_n \otimes P_0^+$ orbits:}
The geometry admits four Killing vectors with periodic generators. The geometry associated with the representative of the orbit in this case has the form \eqref{generaladm} with
\be
F(\rho)=\frac{(\rho^2+\ell^2)^2}{\ell^2\rho^2},\quad N^{\varphi}=\frac{\ell}{\rho^2},\quad \varphi\in[0,2\pi n]\;.
\ee

\item {\bf  $B_0(b) \otimes C(\nu)$ orbits:} The corresponding geometry admits only two Killing vectors $\partial_+$ and $\partial_-$.The metric of the representative element takes general form \eqref{generaladm} with \eqref{mixed-B-C-metric}.

\item {\bf  $B_0(b) \otimes E_n$ orbits:}
This geometry has four periodic Killing vectors, three of them with period $\frac{2\pi}{n}$. The metric for the representative element takes general form \eqref{generaladm} with \eqref{mixed-B-C-metric} with now $L_+=b^2, L_-=-n^2/4$.

\item {\bf  $B_0(b) \otimes P_0^+$ orbits:}
In this case the metric of the representative corresponds to the self dual orbifold of AdS$_3$ \cite{Self-dual-orbifold}. A generic element in this orbit can be obtained as the near horizon limit of geometry corresponding to 
$B_0(b) \otimes B_0(b)$, extremal BTZ black hole orbit \cite{DLCQ}.

\item {\bf  $C(\nu) \otimes E_n$ orbits:}
The corresponding geometry admits four Killing vectors, three of them have period $\frac{2\pi}{n}$. The geometries correspond to chiral particles (those with equal mass and spin) on $N$-fold  ($N=lcm(n,[\nu])$ cover of AdS$_3$.

\item {\bf $C(\nu) \otimes P_0^+$ orbits:}
The representative geometry corresponds to chiral particle on AdS$_3$ in Poincar\'e patch.

\end{itemize}


{}

\end{document}